\documentclass[journal,12pt,onecolumn,draftclsnofoot]{IEEEtran}

\usepackage{cite}
\usepackage[font=footnotesize,caption=false,subrefformat=parens,labelformat=parens]{subfig}
\usepackage{amsmath,amssymb,amsfonts}
\usepackage{bm}
\usepackage{algorithmic}
\usepackage{graphicx}
\usepackage{bm}
\usepackage{multirow}
\usepackage{textcomp}
\usepackage{xcolor}
\usepackage{blindtext}
\def\BibTeX{{\rm B\kern-.05em{\sc i\kern-.025em b}\kern-.08em
    T\kern-.1667em\lower.7ex\hbox{E}\kern-.125emX}}

\usepackage{bbm}    
\graphicspath{ {figures/} }    

 \renewcommand{\baselinestretch}{0.98}
\begin{document}

\title{EMR: A New Metric to Assess the Resilience of Directional mmWave Channels to Blockages}

\author{Fatih Erden,
        Ozgur Ozdemir,
        Ismail Guvenc,~and David W. Matolak
\thanks{This work has been supported in part by NASA under the Federal Award ID number NNX17AJ94A and in part by DOCOMO Innovations, Inc.}%
\thanks{F. Erden, O. Ozdemir, and I. Guvenc are with the Department of Electrical and Computer Engineering, North Carolina State University, Raleigh, NC 27606 (e-mail:\{ferden, oozdemi, iguvenc\}@ncsu.edu).}
\thanks{D. W. Matolak is with the Department of Electrical Engineering, University of South Carolina, Columbia, SC 29208 USA (e-mail: matolak@sc.edu).}
}

\maketitle

\begin{abstract}
Millimeter-wave~(mmWave) communication systems require narrow beams to increase the communication range. If the dominant communication direction is blocked by an obstacle, an alternative and reliable spatial communication path should be quickly identified to maintain connectivity. In this paper, we introduce a new metric to quantify the effective multipath richness (EMR) of a directional communication channel by considering the strength and spatial diversity of the resolved paths, while also taking into account beamwidth and blockage characteristics. The metric is defined as a weighted sum of the number of multipath component~(MPC) clusters, where clustering is performed based on the cosine-distance between the MPCs that have power above a certain threshold. This process returns a single scalar value for a transmitter~(TX)/receiver~(RX) location pair in a given environment. It is also possible to represent the EMR of the whole environment with a probability distribution function of the metric by considering a set of TX/RX locations. Using this proposed metric, one can assess the scattering richness of different communication environments to achieve a particular quality of service~(QoS). 
This metric is especially informative and useful at higher frequencies, such as mmWave and terahertz~(THz), where the propagation path loss and penetration loss are high, and directional non-light-of-sight (NLOS) communication is critical for the success of the network. We evaluate the proposed metric using our channel measurements at 28~GHz in a large indoor environment at a library setting for LOS and NLOS scenarios. \looseness=-1
\end{abstract}


\section{Introduction}
\label{sec:Intro}
High-frequency bands, such as millimeter-wave~(mmWave) and terahertz~(THz), have attracted increasing attention as a solution to the continuously growing data rate demand. Due to very large amounts of available spectrum at these higher frequencies, they have received major attention for 5G, and recently 6G, standardization efforts. For efficient planning of wireless networks, a thorough understanding of the propagation channel characteristics in the respective deployment band is critical. Even though comprehensive knowledge of the sub-6~GHz bands have been acquired already through extensive channel measurements and modeling, there is still much to investigate about mmWave bands for successful deployment and operation of the wireless networks using these bands. 

\begin{figure}[t]
\centering
\includegraphics[trim=1.8cm 0cm 3.2cm 0.5cm, clip,width=0.5\linewidth]{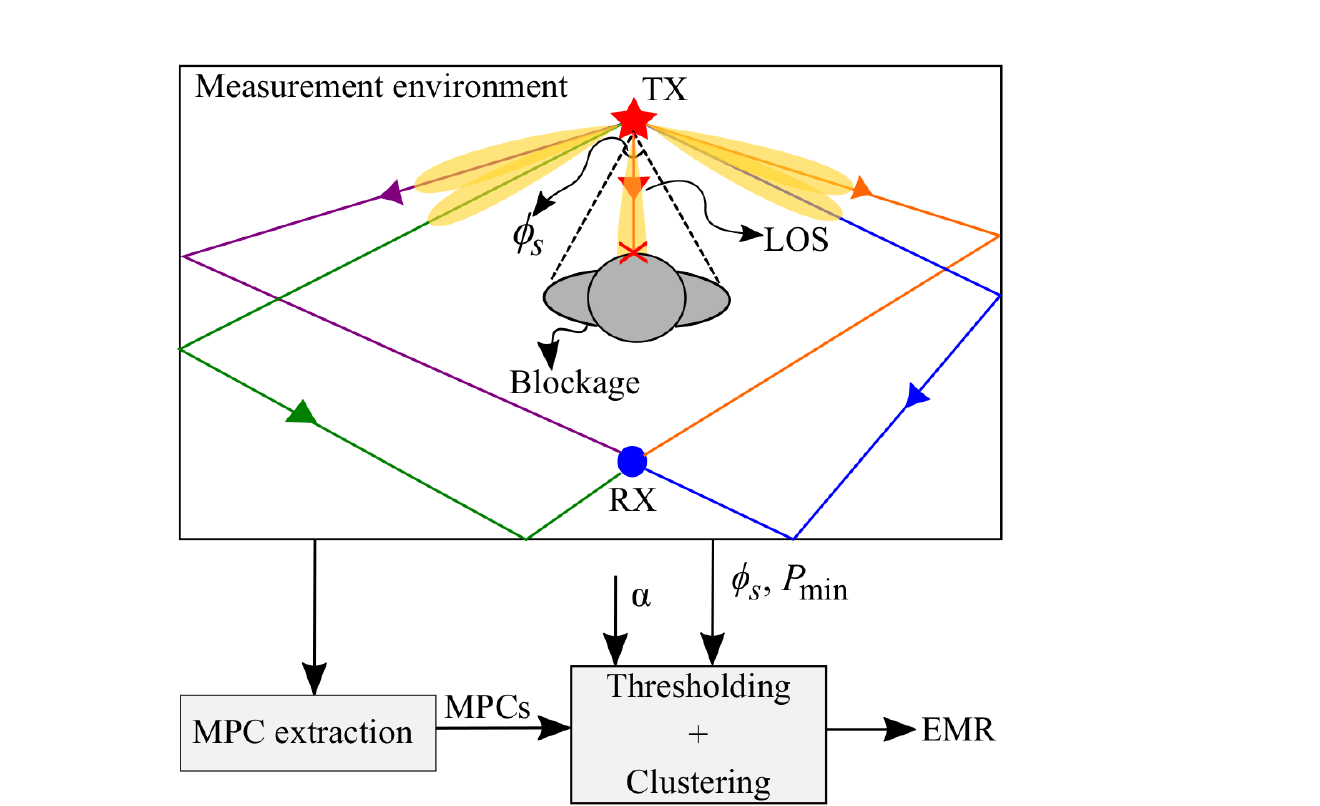}
\caption{An example illustration of the blockage problem and calculation of the EMR metric. There are five paths in total that are above a power level $P_{\min}$ at the RX. The LOS path is blocked and hence the communication should be maintained over the remaining eligible paths. Even though there appears to be four more alternate paths through the reflections from the walls, the two paths on each side can be obscured simultaneously by the same blockage, meaning there are fewer effective backup paths than it appears. The beamwidth is assumed to be smaller than the \emph{blockage width} $\phi_s$.
}
\label{fig:description}
\vspace{-5mm}
\end{figure}

Due to their high frequency, mmWave signals are more sensitive to blockages and attenuate much faster than the sub-6~GHz signals~\cite{Wahab_coverage,Airport31GHz}. For the same reason, free-space path loss of a typical mmWave link is more than an order-of-magnitude larger than that of a traditional sub-6~GHz link~\cite{SouthEastConPaper}. These factors restrict the number of dominant paths (over which communication can be reasonably achieved within QoS constraints) at mmWave frequencies to only a few~\cite{Ko2017}, i.e., line-of-sight~(LOS) path, if available, and a few additional paths through strong reflections from obstacles. The number of dominant paths may further decrease, or there may be none at all, if one or more paths are blocked due to mobile obstacles, as illustrated in Fig.~\ref{fig:description}. However, it is still possible to establish multi-gigabit links between the user and the serving base station~(BS)/access point~(AP) provided that the user location has the minimum necessary received signal strength~(RSS) over at least one path. The most popular solution for compensating the additional loss at high frequencies is to use phased array antennas. Phased array antennas can electrically create and steer beams in different directions so that alternate paths (with more gain due to beamforming) can be utilized in case the most preferable path(s) is (are) blocked~\cite{SouthEastConPaper}. 

Given the foregoing background, the number of backup paths (i.e., dominant paths that can be utilized when more preferable paths are blocked) will increase the chance of establishing and maintaining communication between the users and the BSs/APs. Accurate knowledge of these paths under varying channel conditions is important and can help to evaluate the suitability of an environment for a particular deployment band or to determine the locations of BSs/APs in a given environment to maximize a performance metric (e.g., coverage or throughput). To address this need, this paper defines a new metric, called \textit{effective multipath richness}~(EMR). 

Fig.~\ref{fig:description} illustrates the problem and provides an overview of the EMR metric calculation. The EMR metric assesses the \textit{value} of the multipaths in a measured channel by taking into account the RSS over each path and the overall spatial diversity of the paths. To achieve this, first, paths with power above a certain threshold ($P_{\min}$) at the receiver~(RX) are identified and then clustered based on an angular distance metric. The clustering is performed in an iterative manner so that the resulting clusters satisfy a spatial diversity constraint defined by a beam separation threshold ($\phi_s$). Finally, the relative value of each cluster is determined as a function of the cluster power and a decay coefficient $\alpha$ which determines the \emph{importance of weaker paths}. The value sum is output as the EMR for a particular transmitter~(TX)/RX location pair. 

To represent the overall scattering behavior of an environment, the probability distribution function of the EMR can be used, based on measurements performed at different TX/RX locations in that environment. In essence, the EMR estimates the number of \textit{useful} paths for communication, considering the environment-specific parameters (e.g., blockage size or probability). It also characterizes the gains that can be obtained via beamforming in case the dominant paths get blocked. It will complement the other well-known metrics, such as the root-mean-square~(RMS) delay spread~(DS) or angular spread~(AS), to characterize the channel, and accordingly, to plan the deployment and operation of a mmWave network.\looseness=-1

The rest of the paper is organized as follows. Section~\ref{sec:OtherMetrics} reviews a few well-known metrics that can be used to assess wireless propagation characteristics of an environment and motivates the need for the proposed EMR metric. Section~\ref{Sec:ProposedMetric} introduces the steps to calculate EMR. Section~\ref{Sec:NumericalResults} presents the numerical results on the EMR metric using our data from a 28~GHz channel measurement campaign in an indoor library environment, and Section~\ref{sec:Conclusion} provides concluding remarks.
\section{Characterizing the Channel Behavior}
\label{sec:OtherMetrics}


Information provided by the power delay profile~(PDP) may be mostly sufficient for characterizing the channel at sub-6~GHz frequencies. However, since the angle of arrival/departure (AoA/AoD) information of the multipath components~(MPCs) are of critical importance at high frequencies (i.e., due to beamforming), power angular-delay profiles~(PADPs) of the channel should be extracted at these frequency bands. The PADP can be expressed as~\cite{erden2019}: 
\begin{align}
\label{PADP:eq}
PADP\left(\tau, \bm{\theta}^{\mathrm{AoD}} , \bm{\theta}^{\mathrm{AoA}}\right) = &\sum_{n=1}^{N}  \alpha_n \delta \left(\bm{\theta}^{\mathrm{AoD}}-\bm{\theta}^{\mathrm{AoD}}_n\right)  \\ \nonumber &\times  \delta \left(\bm{\theta}^{\mathrm{AoA}}-\bm{\theta}^{\mathrm{AoA}}_n\right) \delta(\tau-\tau_{n}),
\end{align}
where $N$ is the number of MPCs, $\alpha_n$ is the path gain, $\bm{\theta}^{\mathrm{AoD}}_n = \left[\theta^{\mathrm{AoD,Az}}_n \,\, \theta^{\mathrm{AoD,El}}_n\right]^\intercal$ and $\bm{\theta}^{\mathrm{AoA}}_n = \left[\theta^{\mathrm{AoA,Az}}_n \,\, \theta^{\mathrm{AoA,El}}_n\right]^\intercal$ are the two-dimensional AoD and AoA of the $n$-th MPC in the azimuth and elevation planes, respectively, and $\tau_n$ is the delay of the $n$-th MPC. These parameters can be obtained from channel measurements using appropriate channel sounders or estimated from ray tracing simulations along with accompanying MPC extraction algorithms (e.g., peak search algorithm~\cite{erden2019}, or super-resolution techniques~\cite{Bas2020,Camillo2016}). The relationship between the received power over each path and the corresponding path gain can be written using the link budget as
\begin{equation}
\label{eq:linkbdgt}
P_n=P_{\mathrm{TX}}+G_{\mathrm{TX}}+G_{\mathrm{RX}}+\alpha_n\:[\mathrm{dB}],\quad \forall n\in[N],
\end{equation}
where $G_{\mathrm{TX}}$ and $G_{\mathrm{RX}}$ are the TX and RX antenna gains, and the TX/RX antennas are assumed to be aligned with the $n$-th MPC's AoD/AoA directions. We note that $\alpha_n$ in~\eqref{PADP:eq} is the linear gain, whereas, in~\eqref{eq:linkbdgt}, it is in dB.

Besides PADP, it is also common in the literature to represent different characteristics of the channel concisely using various metrics, such as RMS-DS and RMS-AS. Easy-to-interpret metrics can highlight certain aspects of the channel and help in optimizing performance while planning the network. Next, we will briefly review some of these metrics. We will also provide numerical results based on a few representative channel measurements which will motivate the need for our proposed metric. In the rest of the paper, we will assume that the MPCs have already been extracted from channel measurements or ray tracing simulations as discussed above and use the parameters to compute the metrics.
\subsection{Existing Metrics for Channel Characterization}
RMS-DS is a measure of temporal dispersion of the power. It is calculated using the power and delay parameters of the extracted MPCs as follows~\cite{7857002}:

\begin{equation}
    \tau_{\mathrm{rms}}={\sqrt{\frac{\sum_{n=1}^N P_n(\tau_n-\tau_{\rm avg})^2}{\sum_{n=1}^N P_n}}},
\end{equation}
where $P_n$ is the power of the $n$-th MPC, and $\tau_{\rm avg}$ is the mean delay given by
\begin{equation}
    \tau_{\mathrm{avg}}=\frac{\sum_{n=1}^N P_n\tau_n}{\sum_{n=1}^N P_n}.
\end{equation}

RMS-AS indicates the dispersion of power in the spatial domain. It can be calculated similarly to the RMS-DS as follows~\cite{Yu2005}:
\begin{equation}\label{eq:RMS_AS}
\sigma_{\textrm{rms}}=\sqrt{\frac{\int_{-\pi}^{\pi}P(\omega) (\omega-\omega_\textrm{avg})^{2} d \omega}{\int_{-\pi}^{\pi} P(\omega) d \omega}},
\end{equation}
where $\omega$ is the AoA either in the azimuth or the elevation plane, and $P(\omega)$ is the sum of linear power of the MPCs whose AoA is $\omega$ in the azimuth or the elevation plane. The term $\omega_{\textrm{avg}}$ is the average AoA and given by
\begin{equation}
\omega_{\textrm{avg}}=\frac{\int_{-\pi}^{\pi} P(\omega) \omega d \omega}{\int_{-\pi}^{\pi} P(\omega) d \omega}.
\end{equation}

Another relevant metric to understand the scattering characteristics of the environments is the angular spread coverage~(ASC) introduced in~\cite{Yang2018}. This metric is an extension of the AS metric that was originally proposed in~\cite{durgin1998basic} and quantifies the spatial diversity of the multipaths by 
\begin{equation}
{\textrm{ASC}}=\frac{P_{{\max}}+\sqrt{|F_0|^2-|F_1|^2}}{P_{{\max}}},\label{eq:ASC}
\end{equation}
where $P_{\max}$ is the power of the MPC with maximum power, and
\begin{equation}
F_q=\int_{-\pi}^{\pi} P(\omega) \exp(\mathrm{j}qw) d \omega
\end{equation}
is the $q$-th complex Fourier coefficient of the power arriving at the azimuth angle $\omega$. The original version of the metric does not have the $P_{\max}$ terms in the numerator and the denominator in~\eqref{eq:ASC} and hence returns zero if the available paths have the same AoA azimuth angle. However, the ASC metric can capture the event that the client is in coverage if there is at least one available path.

Finally, we consider a slightly modified version of the RMS-AS, which we refer to as the non-normalized RMS-AS~(nRMS-AS). As its name signifies, the nRMS-AS is calculated as in~\eqref{eq:RMS_AS} but without normalization, i.e., the integral in the denominator of~\eqref{eq:RMS_AS} is set to unit power. The intuition here is that, due to the high attenuation in mmWave frequencies, actual power of the MPCs is a critical feature, and normalizing weighted powers by the total power may remove the multipath richness-related aspects of the channel we are looking for. Given that, the nRMS-AS can be thought of as the RMS value of the absolute power in the angular domain. A similar procedure can be followed to obtain the non-normalized RMS-DS, but we omit this metric in this study for brevity.

\begin{figure*}[t]
\centering
\subfloat[Scenario 1]{\includegraphics[trim=0.8cm 0cm 0.8cm 0.6cm, clip,width=0.45\linewidth]{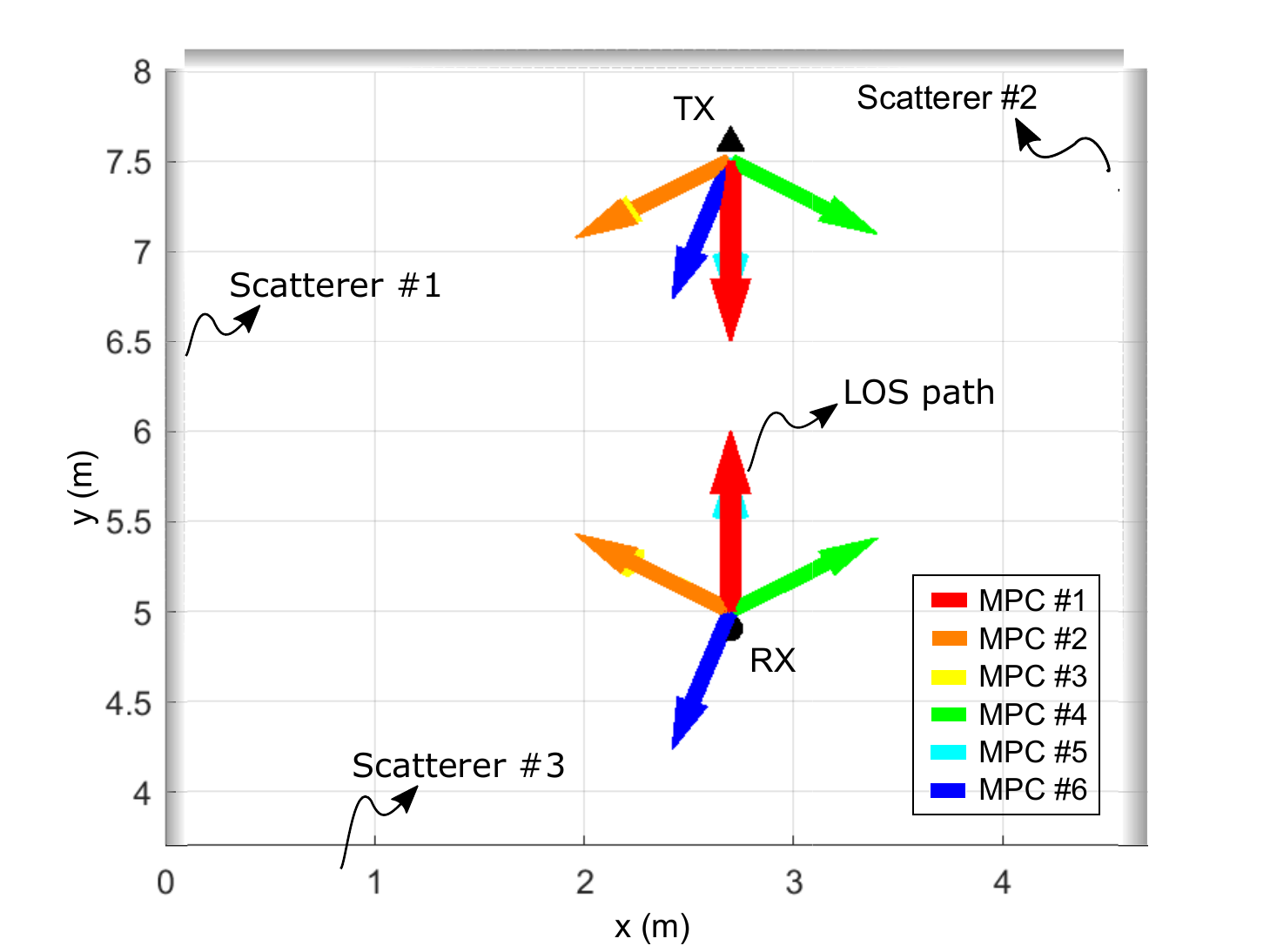}}
\hfill
\subfloat[Scenario 2]{\includegraphics[trim=0.8cm 0cm 0.8cm 0.6cm, clip,width=0.45\linewidth]{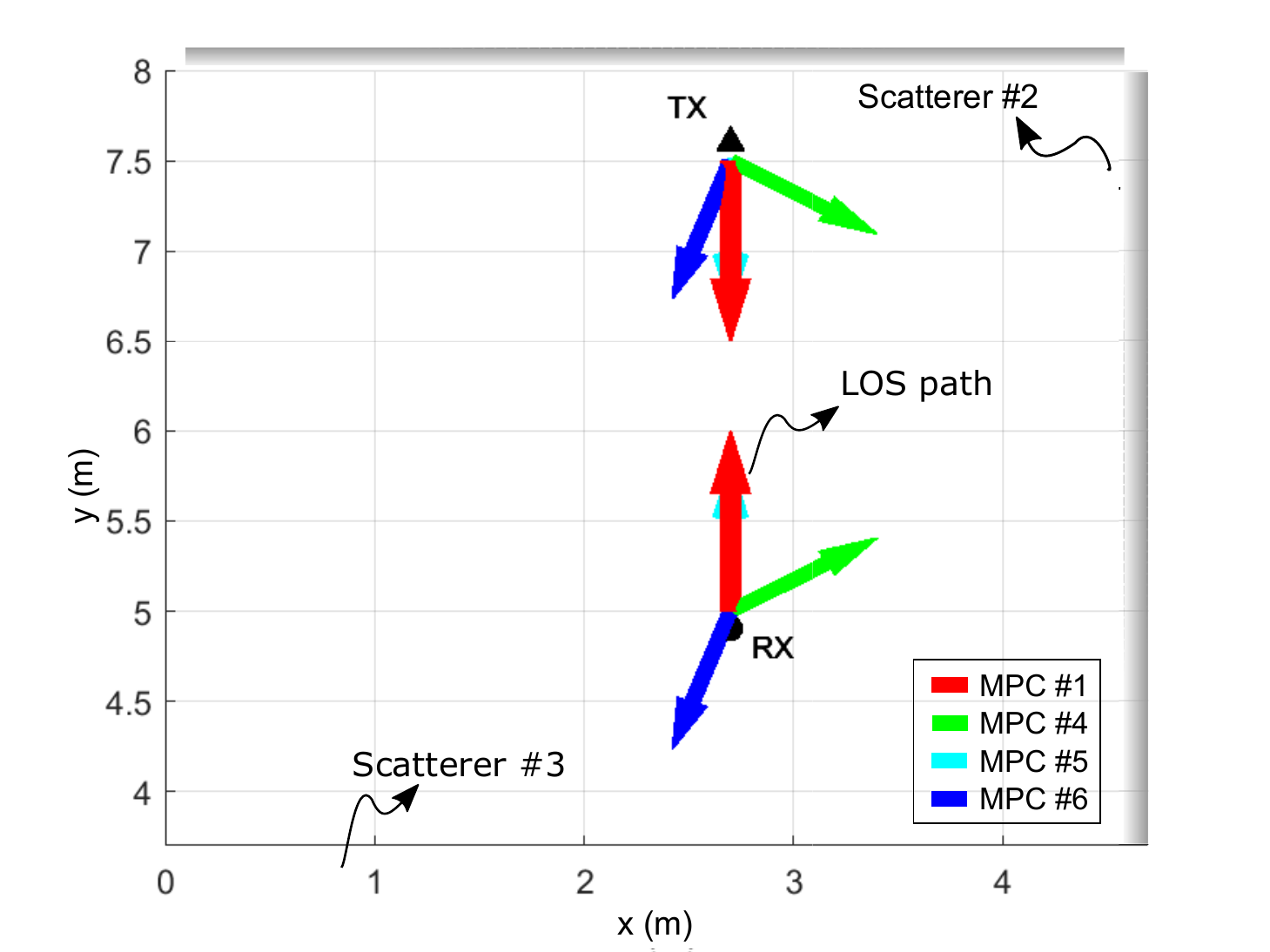}}
\vfill
\subfloat[Scenario 3]{\includegraphics[trim=0.8cm 0cm 0.8cm 0.6cm, clip,width=0.45\linewidth]{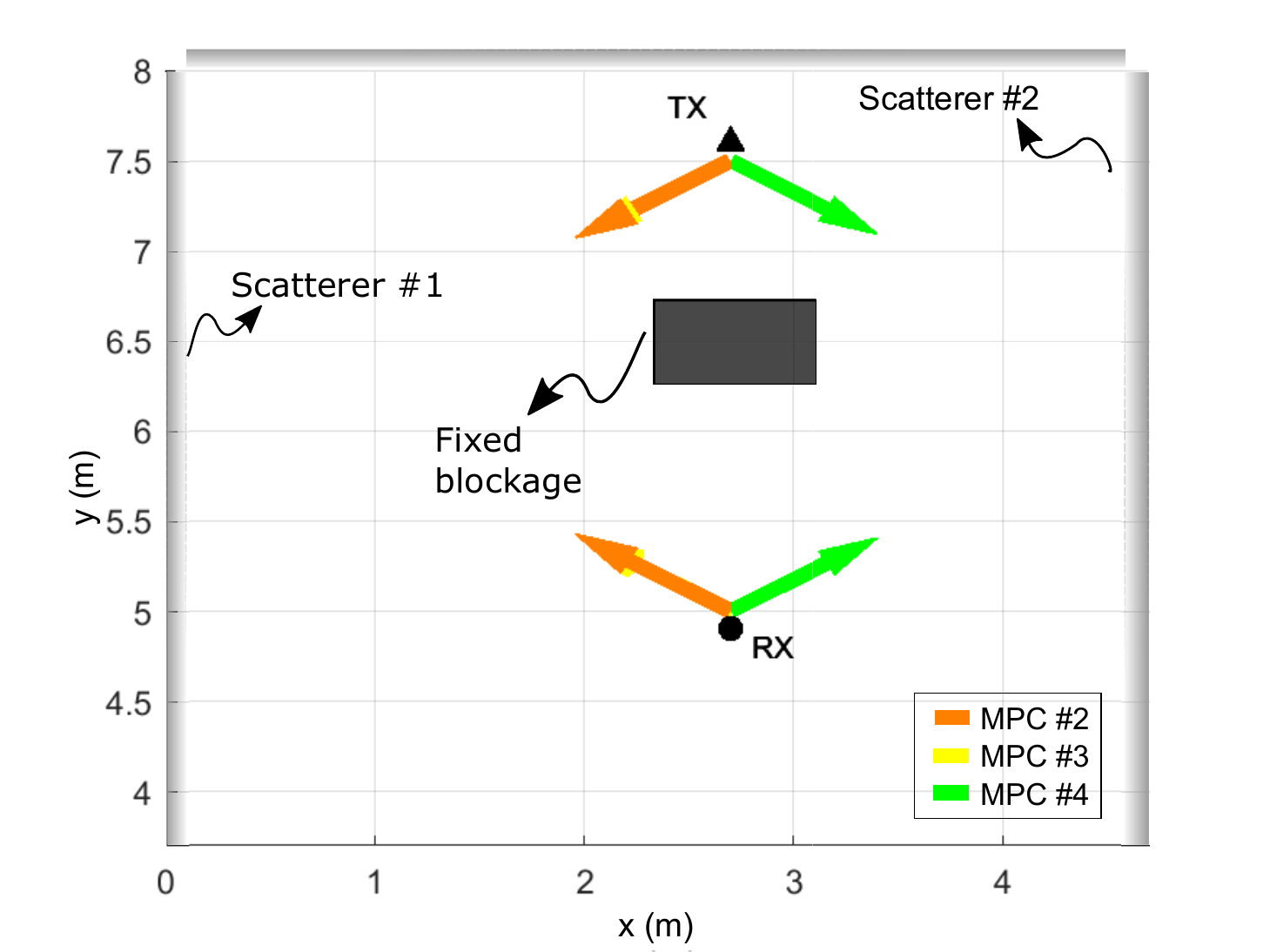}}
\caption{MPCs in the azimuth plane for three different scenarios. Elevation plane is omitted. MPCs are ordered and colored according to their power $P_n$. Warmer colors represent higher power. The MPCs with the same color in different scenarios have the same parameters. The MPC set and hence the spatial/temporal diversity of the paths differ in each scenario; however, RMS-DS, RMS-AS, and ASC fail to reflect these differences, and none of them provides information about the number of backup paths. On the other hand, the EMR can differentiate between the scenarios in terms of their true multipath richness and identifies Scenario 1 as the best environment wih an EMR of 2.42, whereas the other two scenarios have smaller (and comparable) EMRs. 
}
\label{fig:sampleEvaluation}
\vspace{-3mm}
\end{figure*}

\begin{table}[t]
\centering
\renewcommand{\arraystretch}{1.2}
\caption {Parameters of the MPCs Shown in Fig.~\ref{fig:sampleEvaluation}(a).}
\label{Tab:SampleMPCs}
\begin{tabular}{lcccccc}
\hline
\begin{tabular}[c]{@{}l@{}}MPC\\\# (i)\end{tabular}& \multicolumn{1}{c}{\begin{tabular}[c]{@{}c@{}}AoD-Az\\ ($^\circ$)\end{tabular}} & \multicolumn{1}{c}{\begin{tabular}[c]{@{}c@{}}AoD-El\\ ($^\circ$)\end{tabular}} & \multicolumn{1}{c}{\begin{tabular}[c]{@{}c@{}}AoA-Az\\ ($^\circ$)\end{tabular}} & \multicolumn{1}{c}{\begin{tabular}[c]{@{}c@{}}AoA-El\\ ($^\circ$)\end{tabular}} & \multicolumn{1}{c}{\begin{tabular}[c]{@{}c@{}}Power\\ (dBm)\end{tabular}} & \multicolumn{1}{c}{\begin{tabular}[c]{@{}c@{}}Delay\\ (ns)\end{tabular}}  \\
\hline
1  & -90     & 0       & 90     & 0       & -42.00   & 6.51\\
2  & -150     & 0     & 150     & 0       & -53.02   & 13.67\\
3  & -150     & 0     & 150     & -20     & -57.40   & 11.72\\
4  & -30    & 0       & 30     & 0       & -57.67   & 19.53\\
5  & -90      & -20       & 90     & -20       & -58.74   & 7.16\\
6  & -110     & 0       & -110    & 0       & -60.46    & 24.74\\
\hline  
\end{tabular}
\end{table}

\subsection{Evaluation of the Metrics}
This section compares the above metrics evaluated for three scenarios, where the number of extracted MPCs and their spatial distribution vary between the scenarios. Fig.~\ref{fig:sampleEvaluation}(a) shows the first six strongest MPCs (both at the TX and the RX sides) extracted from a sample channel measurement at 28~GHz. The measurement was conducted in a typical office room of size $\approx$ 5~m $\times$ 8~m and equipped with furniture and computers. The highest power is received through the LOS path and denoted by the red arrow. Other paths are through first-order reflections from either the surrounding walls and furniture or the ground, and their power levels are determined by the length of the paths and the material type of the reflectors. It is worthwhile to note that there is no second or higher-order reflections in the list due to severe attenuation at 28~GHz. Parameters of the MPCs are provided in Table~\ref{Tab:SampleMPCs}. This scenario (Scenario 1) is modified by assuming that the Scatterer \#1 is not a good reflector, and thus paths 2 and 3 are not available. It is also assumed that the remaining paths have the same parameters as in Scenario~1. This scenario is referred to as Scenario~2 and illustrated in Fig.~\ref{fig:sampleEvaluation}(b). Lastly, Scenario 1 is also modified such that there is a blockage between the TX and the RX, which obscures the LOS path, the ground-reflected path, and the path reflected from Scatterer~\#3. Therefore, there are only three paths (with the same parameters) as shown in Fig.~\ref{fig:sampleEvaluation}(c). This scenario is referred to as Scenario~3.  

\begin{table}[t]
\centering
\renewcommand{\arraystretch}{1.2}
\caption {Comparison of the EMR with the Other Metrics for the Three Scenarios in Fig.~\ref{fig:sampleEvaluation}.}
\label{Tab:SampleAS_DS}
{\begin{tabular}{lccccc}
\hline
&RMS-DS & RMS-AS & nRMS-AS
& \multirow{ 2}{*}{ASC} &
\multirow{ 2}{*}{EMR}
\\
&(ns)& ($^\circ$) & ($^\circ$)&&
\\
\hline
Scenario~1  &  3.29    & 26.28 & 0.23 & 1.46 & 2.42 \\
Scenario~2  &  2.91    & 20.87 & 0.17 & 1.29 & 1.64 \\
Scenario~3  &  2.67    & 48.05 & 0.14 & 2.18 & 1.81 \\
\hline  
\end{tabular}}
\vspace{-4mm}
\end{table}


The value of the metrics for the above scenarios are given in Table~\ref{Tab:SampleAS_DS}. RMS-AS, nRMS-AS, and ASC metrics are calculated based on the azimuth angles. Since it is a small indoor environment and the measurement was taken at a high frequency (i.e., 28~GHz), RMS-DS values are also small. Moreover, RMS-DS values are close to each other for all three cases. The number of available MPCs and hence the overall spatial distribution vary depending on the environment (e.g., blockage locations); however, RMS-DS does not change significantly, as it represents the delay spread around the mean delay. 

Similarly to the RMS-DS, although the first two strongest MPCs following the LOS path are missing in Scenario~2, there is no much difference between the RMS-AS values of Scenario~1 and Scenario-2. Due to the blockage in Scenario~3, three paths, including the LOS path, are obscured. Therefore, power levels of the remaining MPCs become more comparable (see Table~\ref{Tab:SampleMPCs}). Besides, when the three MPCs are removed, the remaining MPCs become more spread in the angular domain. As a result, a notable increase is observed in the RMS-AS in the last scenario. Similar to the RMS-AS, the ASC metric returns the maximum value for the last scenario. Although the multipath richness is the highest in the first scenario, the first two scenarios have close ASC values. On the other hand, the nRMS-AS is observed to better reveal the channel characteristics we seek out. That is, the nRMS-AS is the highest for Scenario~1, where the spatial diversity is the largest among the other scenarios. Also, the nRMS-AS values are very close to each other for Scenario~2 and Scenario~3, indicating that the two channels are similar to each other in terms of the spatial diversity of the MPCs. However, since the nRMS-AS is a non-normalized metric, its value depends on the absolute power of the MPCs, making it inappropriate to be used to compare the channels where there are large differences in the MPC power levels.

Despite bringing valuable insights into power dispersion in spatial/temporal domains, none of the above metrics provides information about the number of alternate paths and how valuable the paths are when a particular environment is considered. As discussed above, the metrics may return similar values for different sets of MPCs. In addition, relying only on these metrics, it is not straightforward to estimate the channel behavior against different sizes of blockages and the likelihood of a user being in outage. Therefore, we propose the EMR metric, which takes into account the power level of the backup paths and their angular separation, and assesses the relative value of each individual path with respect to the strongest path. The EMR values for the three scenarios are provided in Table~\ref{Tab:SampleAS_DS} and will be interpreted in Section~\ref{Sec:NumericalResults}. 

We note that the AS needs to be calculated separately for azimuth and elevation planes, whereas the EMR metric is calculated based on the three-dimensional angular distance between the paths. However, since the environmental parameters, such as user density or the blockage size/distribution, may differ in different parts of the environments, the EMR should be computed for both the AoA and the AoD for a more complete characterization of the channel. We explain the procedure to calculate the EMR in the following section for only the AoD (i.e., at the TX side). The procedure for the AoA is the same except that, while clustering the MPCs, angular distance between the paths are calculated based on the AoA instead of AoD. \looseness=-1

\section{Effective Multipath Richness~(EMR)}
\label{Sec:ProposedMetric}
To calculate the EMR, two user-defined inputs are required: a minimum power level (at the RX) for the MPCs, and a beam separation threshold. The first input is used to select the MPCs through which the communication can be maintained, whereas the second one is used in clustering the MPCs to estimate the number of backup paths that are \textit{well-spread} over the angular space. Following the power thresholding and clustering processes, the EMR is expressed as the sum of the weighted number of clusters, and the effective richness of the channel in terms of useful paths is revealed. The procedure to find the EMR is explained in detail next. The outcomes of the intermediate steps are demonstrated using a sample measurement (performed in a library environment for TX1-RX1 pair shown in Fig.~\ref{Fig:Meas_Env}(a)). 

\subsection{Thresholding with the Minimum Desired Power Level}
As beamforming will be used at higher frequencies, it would be misleading to include the low-power paths in the calculation of the EMR because these paths will not be of any (or only be a little) help to the communication performance. Therefore, after the extraction of the MPCs, the first step is to identify the MPCs that are above a minimum received power level. This power level should be determined based on the application-specific communication needs, operating frequency, and the available environmental parameters, such as the blockage size/probability statistics. 
For example, in~\cite{8369007}, where a learning-assisted beam search scheme is proposed for indoor mmWave networks at 60~GHz, the minimum required RSS is defined as $-$60~dBm; otherwise the user equipment~(UE) is triggered to find a stronger link. On the other hand, the RSS threshold is considered to be $-84$~dBm for a reliable communication at 60~GHz between the vehicles and the road side units~(RSUs) in an urban environment from Manhattan, NY.

Let $\mathbf{X}=\{\chi_1,\chi_2,\dots,\chi_N\}$ be the set of all MPCs with $N$ being the total number of MPCs. Each MPC in $\mathbf{X}$ can be characterized by six parameters as follows:
\begin{align}
\chi_n=\big\{\theta_{n}^{\mathrm{AoD}, \mathrm{Az}},  \theta_{n}^{\mathrm{AoD},  \mathrm{El}},  \theta_{n}^{\mathrm{AoA}, \mathrm{Az}}, & \theta_{n}^{\mathrm{AoA,El}}, \tau_n, P_{n}\big\}, 
\end{align}
where $n=1, \ldots, N$. Then, the subset of MPCs whose power are above a predetermined level $P_{\min}$ can be defined as 
\begin{equation}
\overline{\mathbf{X}}=\left\{\chi_{n} |\; P_{n}>P_{\min}, \; n=1, \ldots, N\right\}.\label{eq:xbar}
\end{equation}

\begin{figure}[t]
\centering
\subfloat[]{\includegraphics[width=7.5cm]{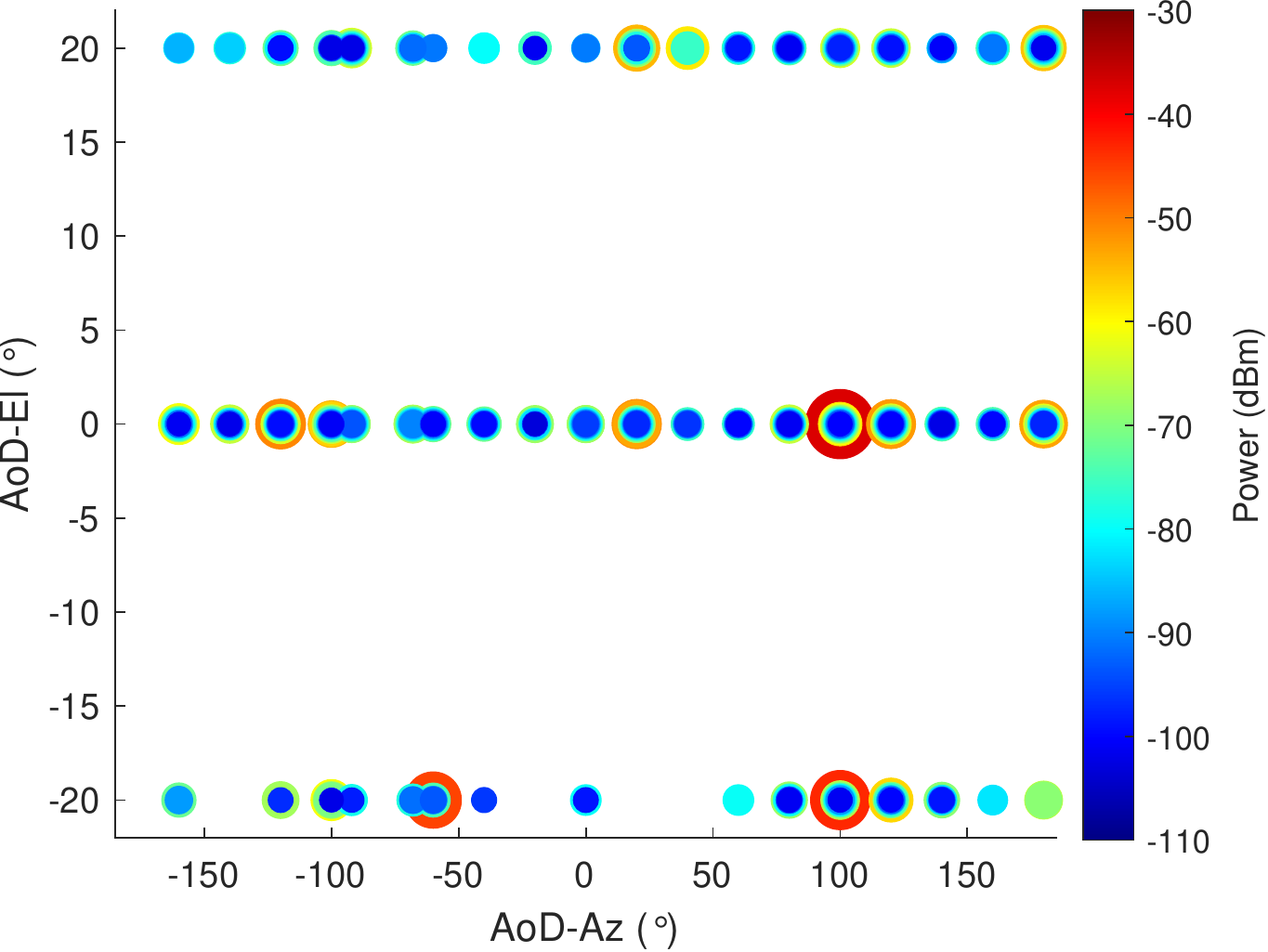}}
\hspace{1cm}
\subfloat[]{\includegraphics[width=7.5cm]{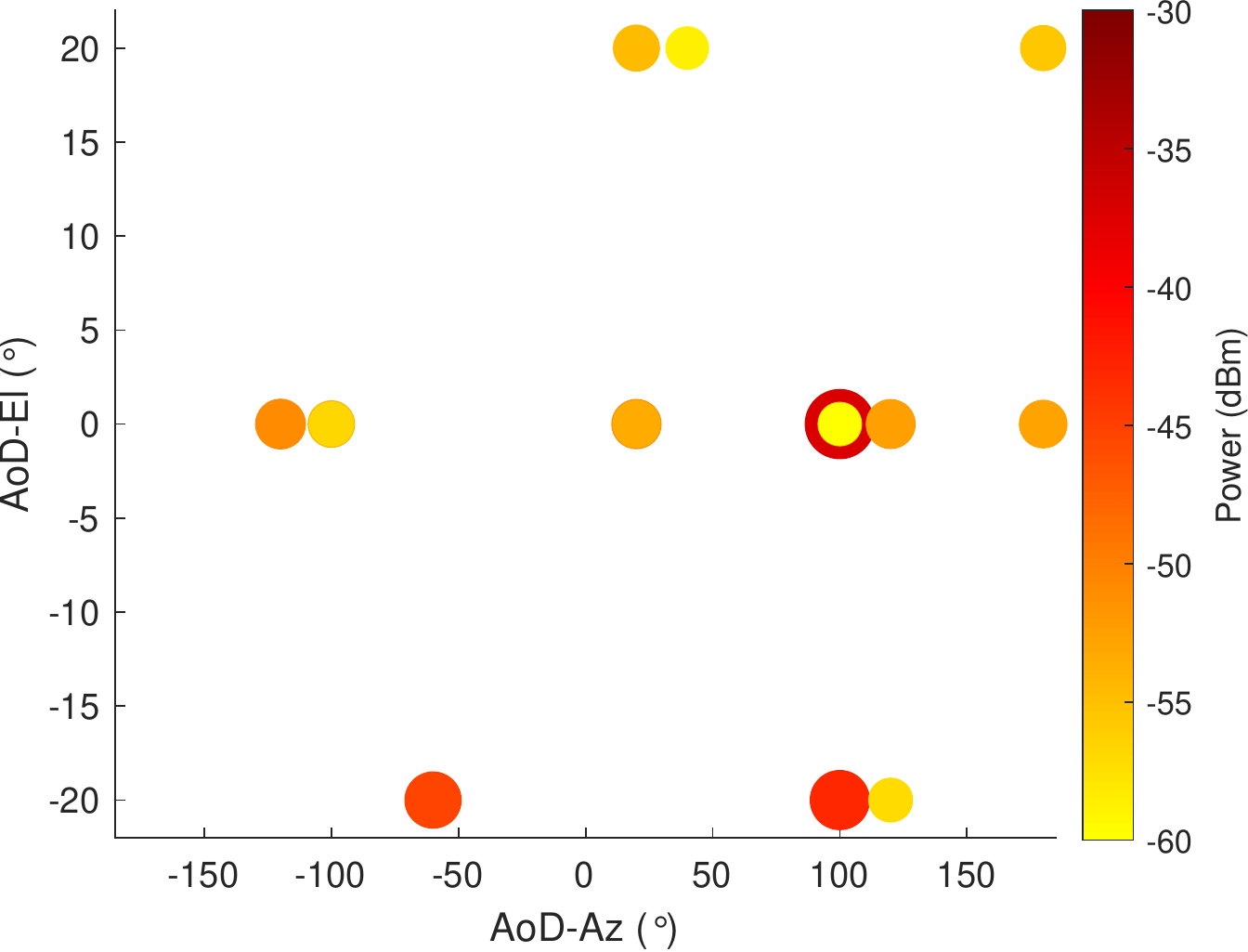}}
\caption{MPCs in the AoD-power space: (a) All MPCs and (b) MPCs after thresholding with $P_{\min}=-60~\textrm{dBm}$. Points with multiple colors indicate multiple MPCs at the same point.}
\label{fig:MPCsAoD_Power}
\vspace{-3mm}
\end{figure}

\begin{table}[t]
\centering
\renewcommand{\arraystretch}{1.3}
\caption {MPCs with Power Level Above $P{\textrm{min}}=-60~\textrm{dBm}$ and Cosine-Distance-Based Clustering Results for $\phi_s=20^{\circ}$.}
\label{Tab:StrongMPCs}
\resizebox{0.485\textwidth}{!}{
\begin{tabular}{lccccccc}
\hline
\begin{tabular}[c]{@{}l@{}}MPC\\\# (i)\end{tabular}& \multicolumn{1}{c}{\begin{tabular}[c]{@{}c@{}}AoD-Az\\ ($^\circ$)\end{tabular}} & \multicolumn{1}{c}{\begin{tabular}[c]{@{}c@{}}AoD-El\\ ($^\circ$)\end{tabular}} & \multicolumn{1}{c}{\begin{tabular}[c]{@{}c@{}}AoA-Az\\ ($^\circ$)\end{tabular}} & \multicolumn{1}{c}{\begin{tabular}[c]{@{}c@{}}AoA-El\\ ($^\circ$)\end{tabular}} & \multicolumn{1}{c}{\begin{tabular}[c]{@{}c@{}}Power\\ (dBm)\end{tabular}} & \multicolumn{1}{c}{\begin{tabular}[c]{@{}c@{}}Delay\\ (ns)\end{tabular}} & \multicolumn{1}{c}{\begin{tabular}[c]{@{}c@{}}Cluster\\\# (i)\end{tabular}} \\
\hline
1  & 100   & 0   & -80  & 0  & -37.33   & 33.85   & 1           \\
2  & 100     & -20     & -80     & 0       & -43.25   & 33.20    & 1    \\
3  & -60     & -20     & -80     & -20     & -45.35   & 42.32   & 2    \\
4  & -120    & 0       & -80     & 0       & -50.75   & 57.94   & 4    \\
5  & 20      & 0       & -20     & 0       & -51.27   & 88.54   & 3    \\
6  & 120     & 0       & -100    & 0       & -51.60    & 37.11   & 1    \\
7  & 120     & 0       & -120    & 0       & -52.35   & 39.06   & 1    \\
8  & 180     & 0       & -140    & 0       & -52.83   & 115.88  & 5    \\
9  & 20      & 0       & -140    & 0       & -53.42   & 102.21  & 3    \\
10 & -100    & 0       & -80     & 0       & -54.08   & 54.68   & 4    \\
11 & 20      & 20      & -20     & 20      & -54.52   & 94.40   & 3    \\
12 & 180     & 20      & -20     & 20      & -55.47   & 107.42  & 5    \\
13 & -100    & 0       & -80     & 0       & -56.60    & 55.99   & 4    \\
14 & 120     & -20     & -100    & 0       & -57.07   & 36.46   & 1    \\
15 & 100     & 0       & -120    & 0       & -57.50    & 38.41   & 1    \\
16 & 40      & 20      & -20     & 20      & -58.69   & 93.74   & 3    \\
17 & 100     & 0       & 120     & -20     & -59.93   & 40.36   & 1
\\ \hline  
\end{tabular}
}\vspace{-3mm}
\end{table}

Fig.~\ref{fig:MPCsAoD_Power}(a) shows all the MPCs (extracted from the sample measurements in Fig.~\ref{Fig:Meas_Env}(a) at TX1-RX1 pair) in the AoD-power space. We note that the number of extracted MPCs depends on several factors, such as multipath delay resolution of the channel sounder or the MPC extraction algorithm. Based on the choice of these factors, one can end up with a much smaller number of MPCs than shown in Fig.~\ref{fig:MPCsAoD_Power}(a). 
Considering the requirements, such as minimum average date rate per user, let us assume the minimum power level $P_{\min}$ for an MPC to be eligible as a backup path (after a blockage, in a different beam direction) to be $-60$~dBm. Then, the number of MPCs reduces to only 17, and the resulting MPCs are shown in Fig.~\ref{fig:MPCsAoD_Power}(b). The parameters of these MPCs are provided in Table~\ref{Tab:StrongMPCs}. As it is clear from Fig.~\ref{fig:MPCsAoD_Power}(b), three-dimensional AoDs for some of the MPCs are very close to each other, which may result in multiple MPCs being blocked in the presence of an obstacle (of certain size) that is at a certain distance to the TX. As a result of this observation, the likelihood of finding a link between the TX and the RX will be higher when the MPCs spread out over the AoD azimuth and AoD elevation space compared to when the MPCs exhibit clusters. Therefore, for a reliable assessment of the multipath richness of a channel, one should also take into consideration the spatial diversity of the MPCs. \looseness = -1

\subsection{Cosine Distance-Based Clustering}\label{Sec:CosDistClustering}
Having identified the paths over which a link can be established, the next step is to find the number of effective alternate paths (i.e., beam directions that are separated by at least a user-defined angular distance from any other). Here we assume that the beamwidth is smaller compared to the blockage angle $\phi_s$, as shown in Fig.~\ref{fig:description}. To achieve higher gains, multiple antenna elements with narrow beams are used at mmWave frequencies. This also limits the number of beam directions. For example, it is shown in~\cite{2018tutorial} that the number of beam directions at the gNB side (with a 8~$\times$~8 array and beamwidth 13$^\circ$) is only 10 in azimuth when the gNB scans a total of 120 degrees.

Unlike the approaches that aim to parameterize the channel impulse response by clustering the MPCs based on AoA, AoD, and delay information (e.g.,~\cite{Huang17,Rappaport15}), we cluster the MPCs only using their angular parameters. This way, it is possible to find how many alternate paths are available if some of the paths are blocked. Omitting the parameters other than the angles in the interested domain (in this case, the AoD), each MPC in $\overline{\mathbf{X}}$ in \eqref{eq:xbar} can be represented by a vector in three-dimensional space as
\begin{equation}
\label{eq:unitVector}
\chi_{l}=\left\{r_l,\theta_{l}^{\mathrm{AoD}, \mathrm{Az}}, \theta_{l}^{\mathrm{AoD}, \mathrm{El}}\right\},
\end{equation}
for $l=1,\ldots,L$, where $L$ is the number of MPCs in $\overline{\mathbf{X}}$, and $r_l$ is the magnitude of the vector $\chi_{l}$. While $r_l$ may be defined as the individual powers of the MPCs, since here only the angles are of interest, $r_l$ can be simply set to 1, so $\chi_l$ becomes a unit vector with the given elevation and azimuth angles. 

For clustering the MPCs in~\eqref{eq:unitVector}, an iterative cosine distance-based k-means clustering algorithm is used. At each iteration $\overline{\mathbf{X}}$ is grouped into a number of clusters that is equal to the current iteration count, and this process is repeated until the desired spatial distance between the cluster centroids is achieved. The cosine distance between any two vectors $a$ and $b$ is defined as follows:
\begin{equation}
d(a,b)=1-\cos(\phi),    
\end{equation}
where $\phi$ is the angle between the vectors in three-dimensional space. If $a$ and $b$ point in the same direction ($\phi=0^{\circ}$), then $d(a,b)=0$, or if they point in opposite directions ($\phi=180^{\circ}$), then the distance attains its maximum, and $d(a,b)=2$. 

Let $\bm{c_k}=\{c_{k,1},\ldots,c_{k,L}\}$ be the set of cluster centroids at $k$-th ($k\leq L$) iteration, where $c_{k,l}$ is a $1\times 3$ vector that represents the centroid of $\chi_{l}\in \overline{\mathbf{X}}$. It should be noted that, although each MPC will be matched to a centroid, there will be $k$ unique clusters and hence centroids. Each centroid is the coordinate-wise mean of the points in a cluster, after normalizing those points to unit Euclidean length. Once the centroids are found for the current number of clusters, the angular distance between any MPC and the centroid of the cluster to which that MPC belongs is calculated. If all the MPCs are at an angular distance from their centroids of less than a beam separation threshold $\phi_s$, i.e.,
\begin{equation}
\label{eq:stopp}
\phi_{k,l}=\angle c_{k,l} \chi_{l}<\phi_{s}, \forall l \in(1, \ldots, L)~,
\end{equation}
then the iterations are terminated, and the current iteration count $k$ is returned as the number of clusters. 

The parameter $\phi_s$ is a design parameter, and it represents the size or the angular width of the blockages that are likely to obscure the possible links between the TX and the RX in a given environment (see Fig.~\ref{fig:description}). When $\phi_s$ is higher, it gets more likely that the paths close to each other will be blocked. As a result, there will be fewer paths over which the signals can be transmitted. We point out that if the angular resolution of the TX/RX antennas is much lower than $\phi_s$, then the clustering process can be skipped. In such a case, each resolved path can be treated as a cluster, and their powers can be directly plugged into~\eqref{eq:scam} to calculate the EMR.

It should be recalled that initializing the centroids at different locations may lead the k-means algorithm to return different clusters. It is also possible that some initializations may result in local optima. Both scenarios introduce a bias in the EMR values. Therefore, at each iteration, we randomly initialize $k$ centroids, run the k-means algorithm, and compute the cost function. This process is repeated $K$ times, and the clustering that yields the lowest cost is picked for the current iteration. For the channel measurements used in this study, we observed that it is sufficient to set $K=20$ for the clusters to converge; however, larger $K$ values may be required if there are too many eligible paths after the thresholding step.
 

\begin{figure}[t!]
\centering
\subfloat[]{\includegraphics[trim=0cm 0cm 1cm 0.7cm, clip,width=0.4\columnwidth]{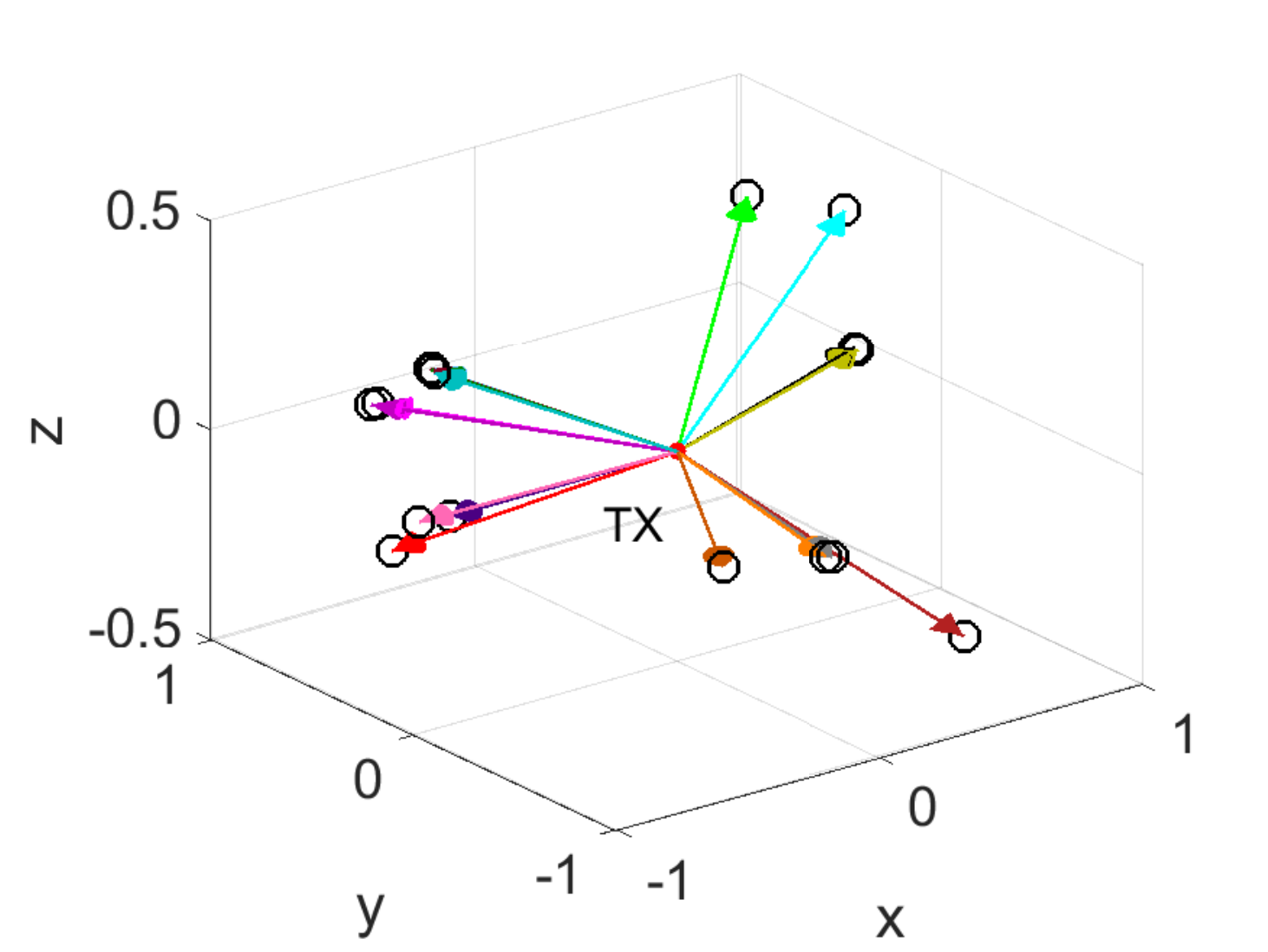}}
\hspace{0.5cm}
\subfloat[]{\includegraphics[trim=0cm 0cm 1cm 0.7cm, clip,width=0.4\columnwidth]{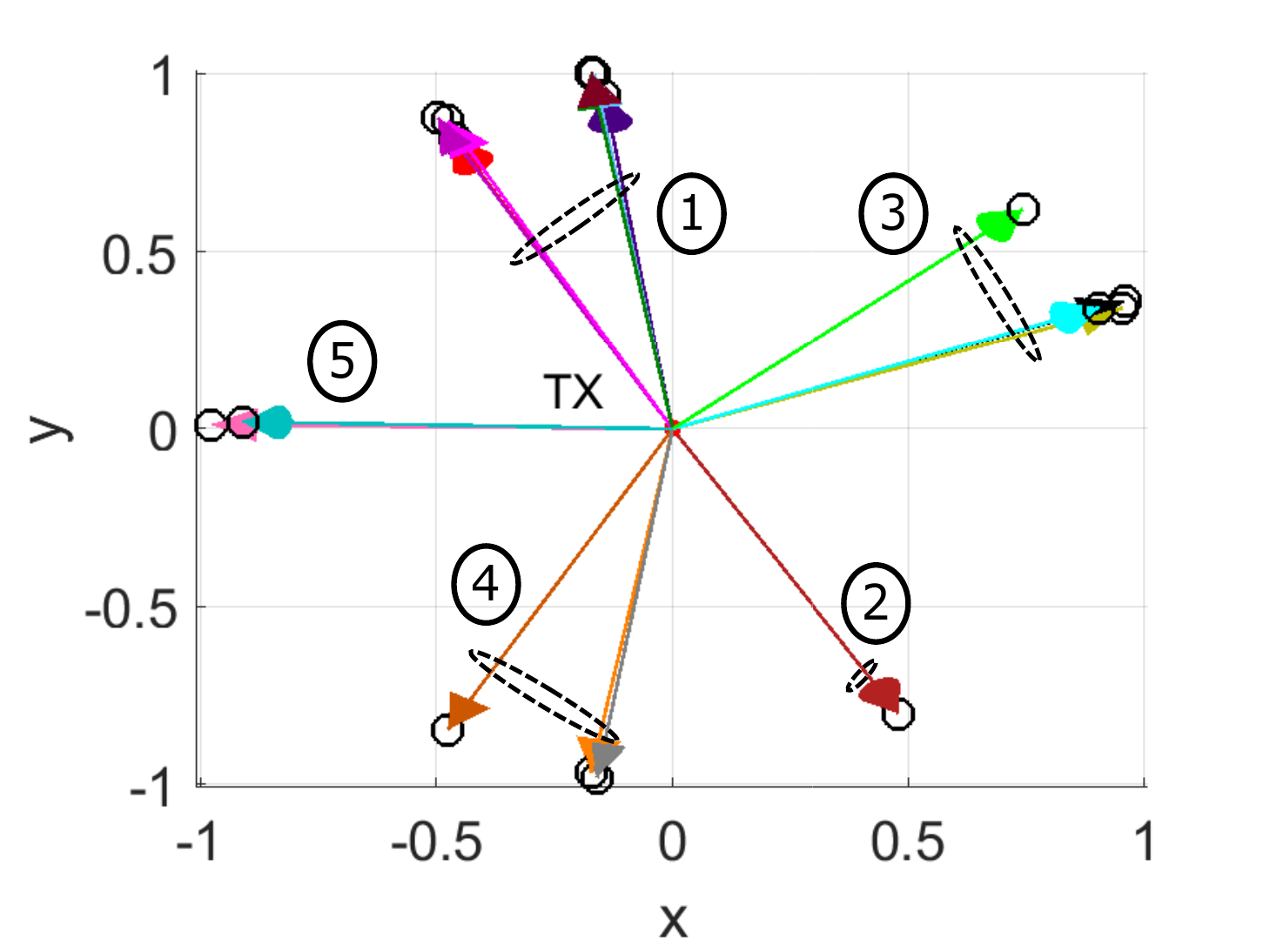}}
\vfill
\subfloat[]{\includegraphics[trim=0cm 0cm 1cm 0.9cm, clip,width=0.4\columnwidth]{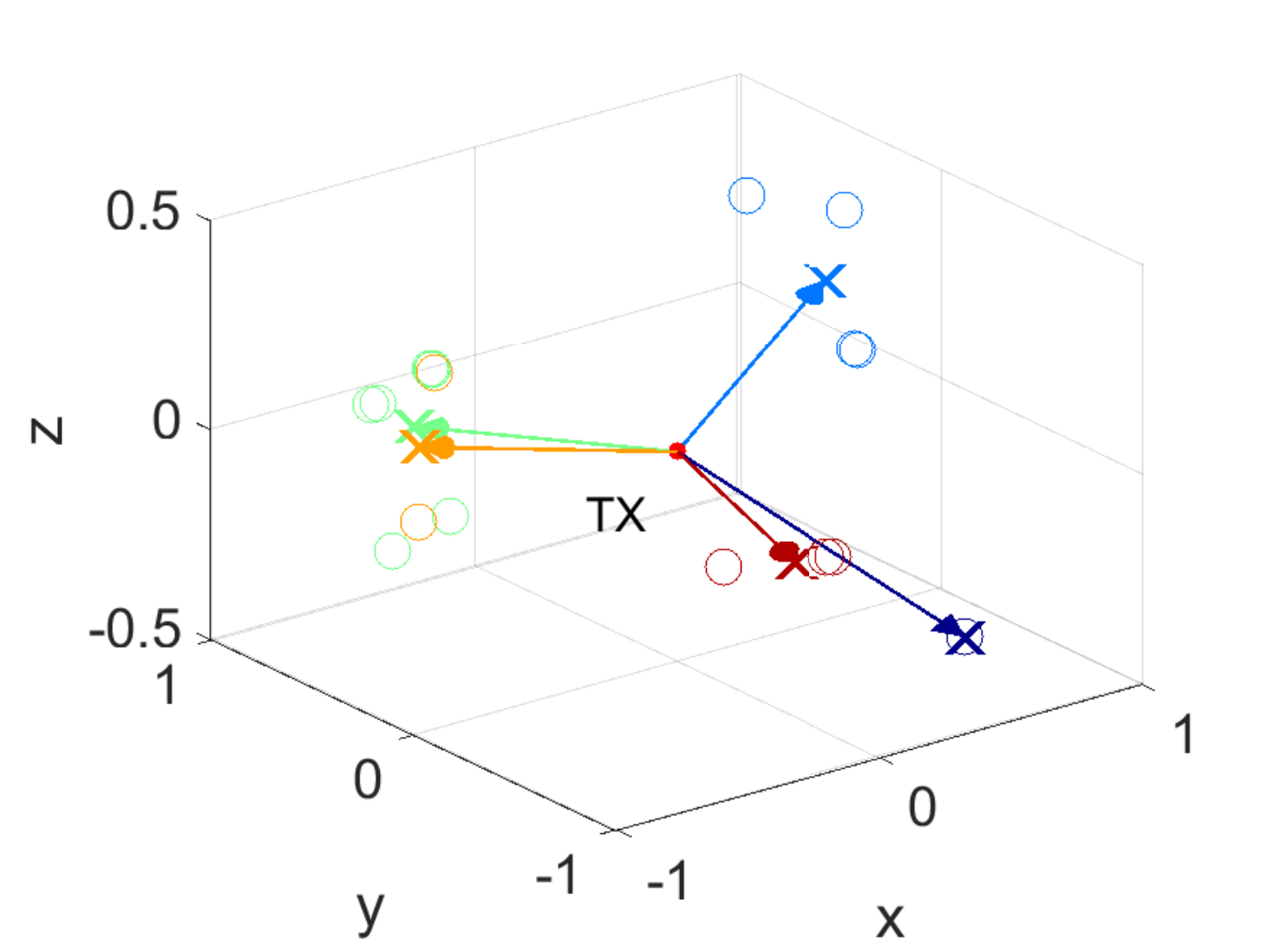}}
\hspace{0.5cm}
\subfloat[]{\includegraphics[trim=0cm 0cm 1cm 0.7cm, clip,width=0.4\columnwidth]{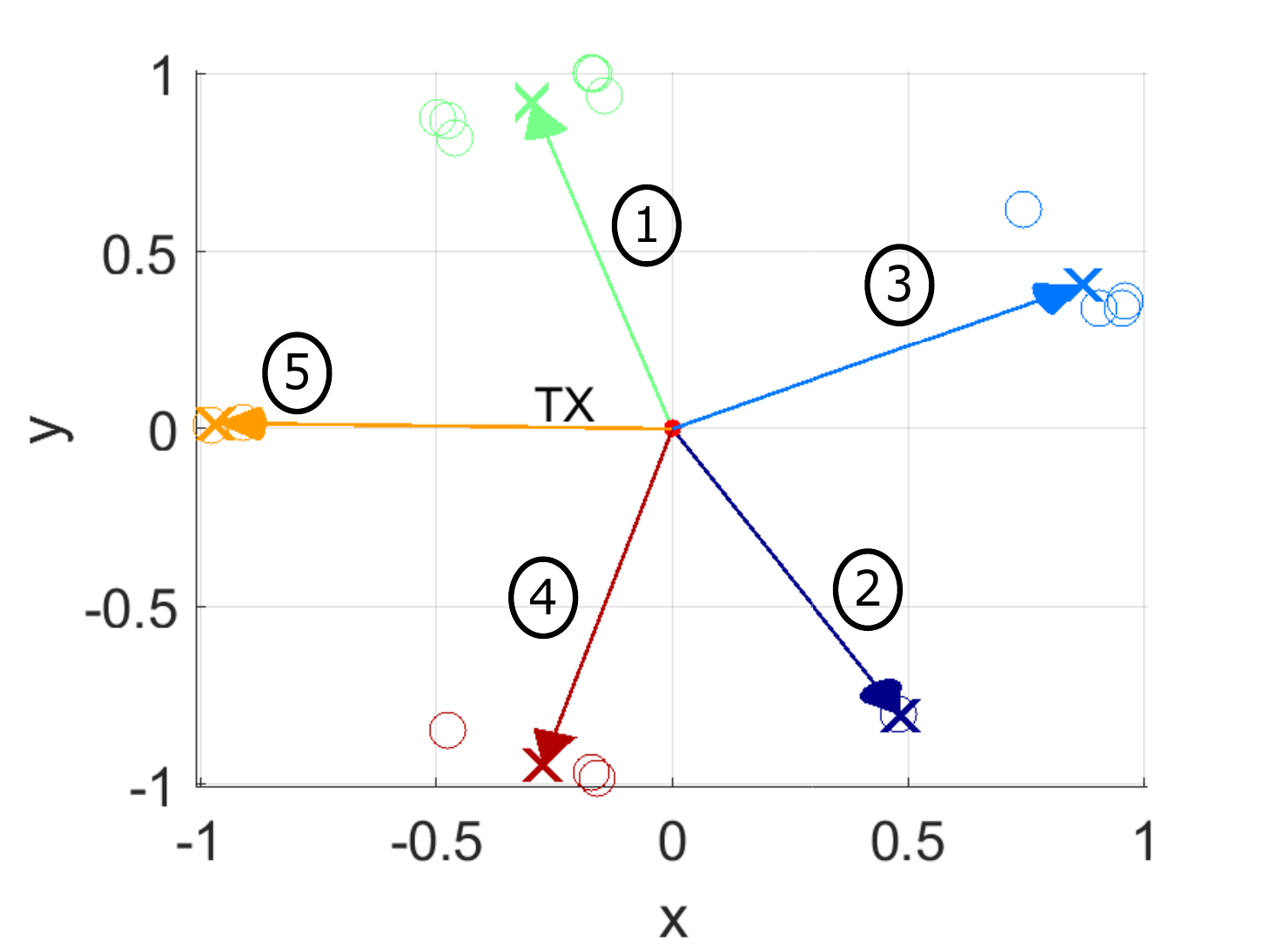}}
\caption{MPCs (in unit vector representation) after thresholding with $P_{\min}=-60~\textrm{dBm}$ (a) in 3D and (b) in 2D. Representation of the MPCs by centroids after our clustering algorithm with $\phi_s=20^{\circ}$ (c) in 3D and (d) in 2D. The centroids after clustering in (d) are more evenly spread out and hence they better capture the number of beam directions compared to (b) for the given $\phi_s$.\looseness = -1}
\label{fig:MPCsArrow}
\vspace{-5mm}
\end{figure}

We applied our clustering algorithm on the MPCs listed in Table~\ref{Tab:StrongMPCs}. The stoppage criterion in~\eqref{eq:stopp} is satisfied after $k=5$ iterations, when $\phi_s$ (characterizing the blockage angle) is set to $20^{\circ}$, and the cluster IDs of the MPCs are given in the last column of the table. Thus, when the resilience of the channel to blockages and hence the spatial diversity of the useful MPCs are of concern, for the given $\phi_s$ it may be presumed that there are only four alternate beam directions outside of the dominant beam direction. The MPCs in unit vector representation after thresholding and clustering steps are shown in 3D view in Fig.~\ref{fig:MPCsArrow}(a) and Fig.~\ref{fig:MPCsArrow}(c), respectively, and in 2D view along with the cluster IDs in Fig.~\ref{fig:MPCsArrow}(b) and Fig.~\ref{fig:MPCsArrow}(d). In Fig.~\ref{fig:MPCsArrow}(a) and Fig.~\ref{fig:MPCsArrow}(b), MPCs are indicated with a circle and the corresponding unit vectors, whereas the cross marks in Fig.~\ref{fig:MPCsArrow}(c) and~Fig.~\ref{fig:MPCsArrow}(d) denote the centroids of the clusters. By comparing Fig.~\ref{fig:MPCsArrow}(b) and Fig.~\ref{fig:MPCsArrow}(d) and treating the centroids as new MPCs, it can be seen that the new MPCs are more evenly spread out and hence better summarize the usable number of backup paths for the given $\phi_s$.

\begin{table}[t!]
\centering
\renewcommand{\arraystretch}{1.3}
\caption {Value of Each Cluster and the EMR for Different $\alpha$ Values ($\phi_s=20^\circ$ and $P{\textrm{min}}=-60~\textrm{dBm}$).}
\label{Tab:ClusterPowervsWeights}
\begin{tabular}{lcccc}
\hline
\multicolumn{1}{l}{\begin{tabular}[l]{@{}l@{}}Cluster\\ \# (i)\end{tabular}} & \multicolumn{1}{c}{\begin{tabular}[c]{@{}c@{}}Cluster power \\ (dBm)\end{tabular}} & \multicolumn{1}{c}{\begin{tabular}[c]{@{}c@{}} \\ ($\alpha$=0.1)\end{tabular}} & \multicolumn{1}{c}{\begin{tabular}[c]{@{}c@{}}Value \\ ($\alpha$=0.4)\end{tabular}} & \multicolumn{1}{c}{\begin{tabular}[c]{@{}c@{}} \\ ($\alpha$=0.7)\end{tabular}} \\
\hline
1  & -36.02   & 1.00   & 1.00  & 1.00   \\
2    & -45.35  & 0.80   & 0.42  & 0.22 \\3   & -47.72    & 0.76  & 0.33 & 0.15  \\4   & -48.39  & 0.75  & 0.31  & 0.13 \\5   & -50.94  & 0.70 & 0.24  & 0.08     \\ \hline\hline
\multicolumn{2}{c} {\bm{$\rho$}}  & \textbf{4.01} & \textbf{2.30}  & \textbf{1.58}\\
\hline
\end{tabular}
\vspace{-5mm}
\end{table}

\subsection{Value of the Backup Paths and the EMR Metric}
The last step in calculating the EMR is to compute the value of the backup paths. We note that, after the clustering process, backup paths refer to the cluster centroids with the difference that the magnitude term $r_l$ in~\eqref{eq:unitVector} is now the sum of the linear power of the MPCs in the corresponding cluster. The value of each backup path depends on its power level relative to the strongest path in the same measurement. Let $P_{c,m}$ be the power of the $m$-th cluster and $K$ be the total number of clusters. Then, the EMR is calculated by summing the value of the individual paths as follows 
\begin{equation}
\label{eq:scam}
    \rho=\sum_{m=1}^K\left(\frac{P_{c,m}-P_{\min}}{P_{c,{\max}}-P_{\min}}\right)^\alpha,
\end{equation}
where $P_{c,\max}$ is the power of the strongest cluster, and $\alpha$ is the decay coefficient that takes values in $[0,1]$. Min-max normalization in~\eqref{eq:scam} ensures that the maximum value of any cluster is limited to 1. On the other hand, $\alpha$ is a tunable parameter that determines the relative value of the clusters with respect to the cluster with the highest total power. If $\alpha=0$, then the value of each cluster will be the same and equal to 1. As $\alpha$ is increased towards 1, the value of the weaker clusters and hence the EMR decreases. It is also important to stress that, in normalizing the cluster powers, $P_{\min}$ is used instead of $P_{c,\min}$. This way, while determining the value of the cluster MPCs, the metric also takes into account how much they are above the minimum required power $P_{\min}$.  

\begin{figure}[!t]
\centering
\subfloat[]{\includegraphics[trim=0cm 0cm 0cm 0cm, clip,width=0.7\linewidth]{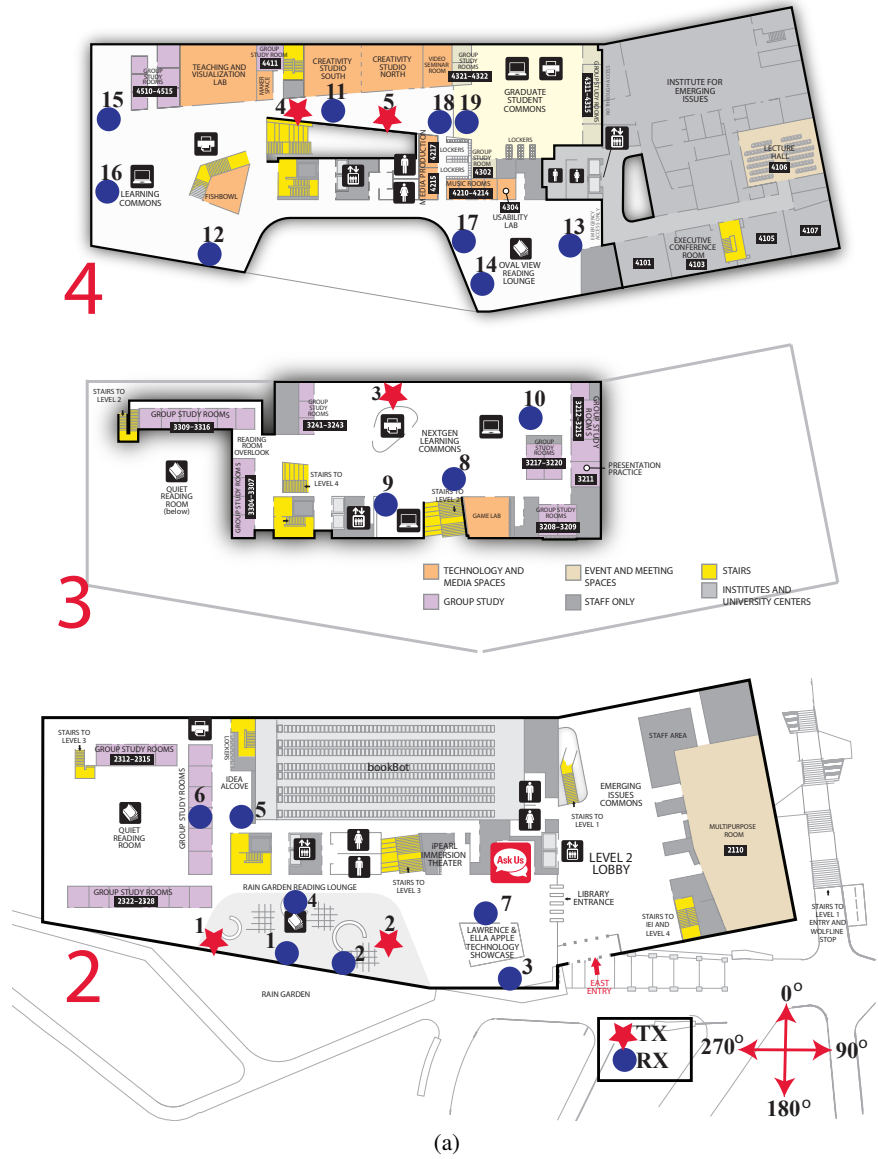}}
\hspace{-5pt}
\subfloat[]{\includegraphics[trim=3cm 3cm 1.5cm 3cm, clip,width=0.6\linewidth]{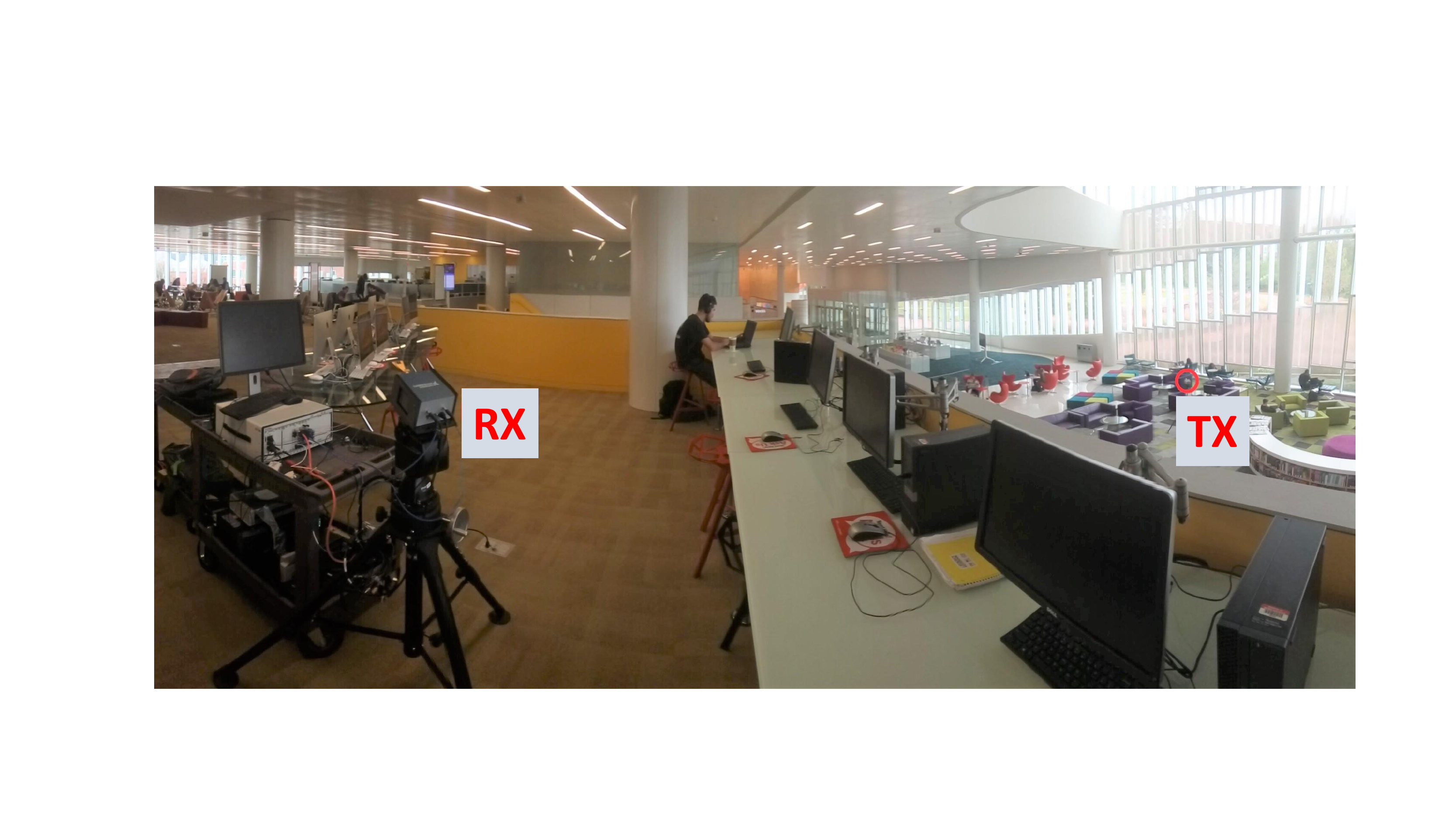}}
\caption{
(a) The Hunt library floor plans with TX/RX locations for a total of 19 LOS and NLOS measurements. 
(b) Snapshot of the measurement scenario for TX1-RX3 pair.}
\label{Fig:Meas_Env}
\end{figure}

Table~\ref{Tab:ClusterPowervsWeights} shows the power and the value of each cluster (i.e., the summand in~\eqref{eq:scam}) or the backup paths, and the EMR for three different $\alpha$ values when $\phi_s=20^{\circ}$ and $P_{\min}=-60$~dBm. The cluster with the highest power, which includes the LOS path for this sample measurement, has a value of 1 independent of the $\alpha$ value. Therefore, if there is at least one path above $P_{\min}$, then the EMR will be greater than or equal to 1. As $\alpha$ is increased from 0.1 to 0.7, the relative value of the weaker clusters decreases in accordance with the above explanation, and the EMR representing this particular TX/RX location decreases from 4.01 to 1.58. So $\alpha$ value can be determined based on how much one wants to rely on the weaker paths to establish communication in a given environment, and the decision on the suitability of a TX location or the whole environment can be made by interpreting the EMR metric.

\section{Numerical Results}
\label{Sec:NumericalResults}
In this section, we evaluate the EMR metric using our channel measurements at the Hunt library at NC State University Centennial Campus for the TX-RX locations shown in Fig.~\ref{Fig:Meas_Env}. Before that, we go back to Table~\ref{Tab:SampleAS_DS} where we compare the EMR metric with the existing metrics for the scenarios shown in Fig.~\ref{fig:sampleEvaluation}. When the TX side is considered with parameters $\{\alpha, \phi_s,P_{\min}\}=\{0.1,20^\circ,-60~\textrm{dBm}\}$, the EMR is equal to 2.42, which means that, in addition to the LOS path, there are \emph{at least} two well-separated paths with power above the required minimum. This is because, from~\eqref{eq:scam}, the value of a backup path can be equal to at most 1 so one can conclude there are at least $\lfloor\rho\rfloor$ backup paths, where $\lfloor\cdot\rfloor$ is the floor function. For example, if the LOS path (MPC \#1) is blocked, then most likely the paths of MPC~\#5 and MPC~\#6 will be blocked as well; however, the EMR indicates that the communication can be maintained over two alternate paths which are separated by more than $20^\circ$ from any other path. Similarly, as it can be seen from Fig.~\ref{fig:sampleEvaluation}(b) and Fig.~\ref{fig:sampleEvaluation}(c), MPCs can be clustered into two in both Scenario~2 and Scenario~3. Therefore, the EMR values are close to 2 in both cases indicating the availability of at least one backup path. The EMR value is slightly larger in Scenario~3 than in Scenario~2 because the total power of the clusters are closer to each other in Scenario~3 than in Scenario~2, increasing the value of the weaker cluster.\looseness=-1

\begin{figure}[b!]
\vspace{-4mm}
\centering
\includegraphics[width=0.5\linewidth]{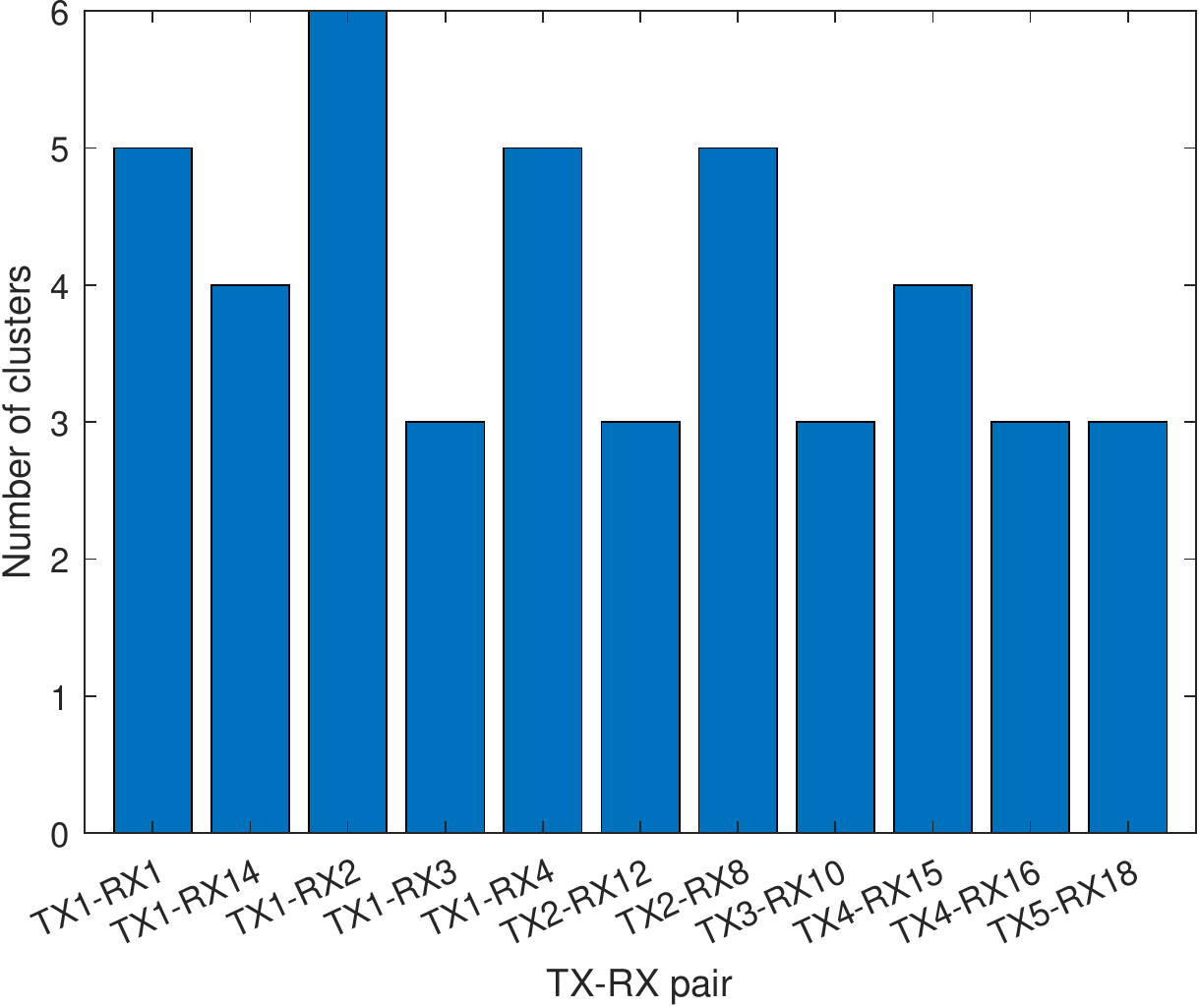}
\caption{Number of clusters for the LOS library measurements using the cosine distance-based clustering in Section~\ref{Sec:CosDistClustering} ($P_{\min}=-60$~dBm, $\phi_s=20^{\circ}$).}
\label{fig:numClusters}
\vspace{-5mm}
\end{figure}

Library measurements include 11 LOS and 8 NLOS scenarios for TX-RX separation distances ranging from 10~m to 50~m. Other details about the measurements and the measurement environment can be found in~\cite{fatih_library}. For the extraction of the MPCs, we used the peak searching algorithm in~\cite{erden2019}. We start with the LOS measurements. Fig.~\ref{fig:numClusters} shows the number of clusters for each TX-RX pair. It is observed that, independent of the TX-RX separation distance (up to 50~m) and height difference (up to two floors), there are at least three well-separated paths that can be utilized. Depending on the surrounding furniture, walls, and electrical appliances, the number of alternate paths increases up to six. Although these observations are of considerable value, they are not sufficient to reveal the true multipath richness of a channel. For example, it is highly likely to have multiple TX locations that will lead to the same number of clusters for the same RX location, and in such a case, the information about the number of clusters does not provide sufficient insight into which TX location is more favorable than others. For this reason we use the EMR metric but we also note interpreting the EMR with the number of clusters can provide additional insights on the propagation channel.\looseness=-1

\begin{figure}[t!]
\centering
\subfloat[]{\includegraphics[width=0.49\columnwidth]{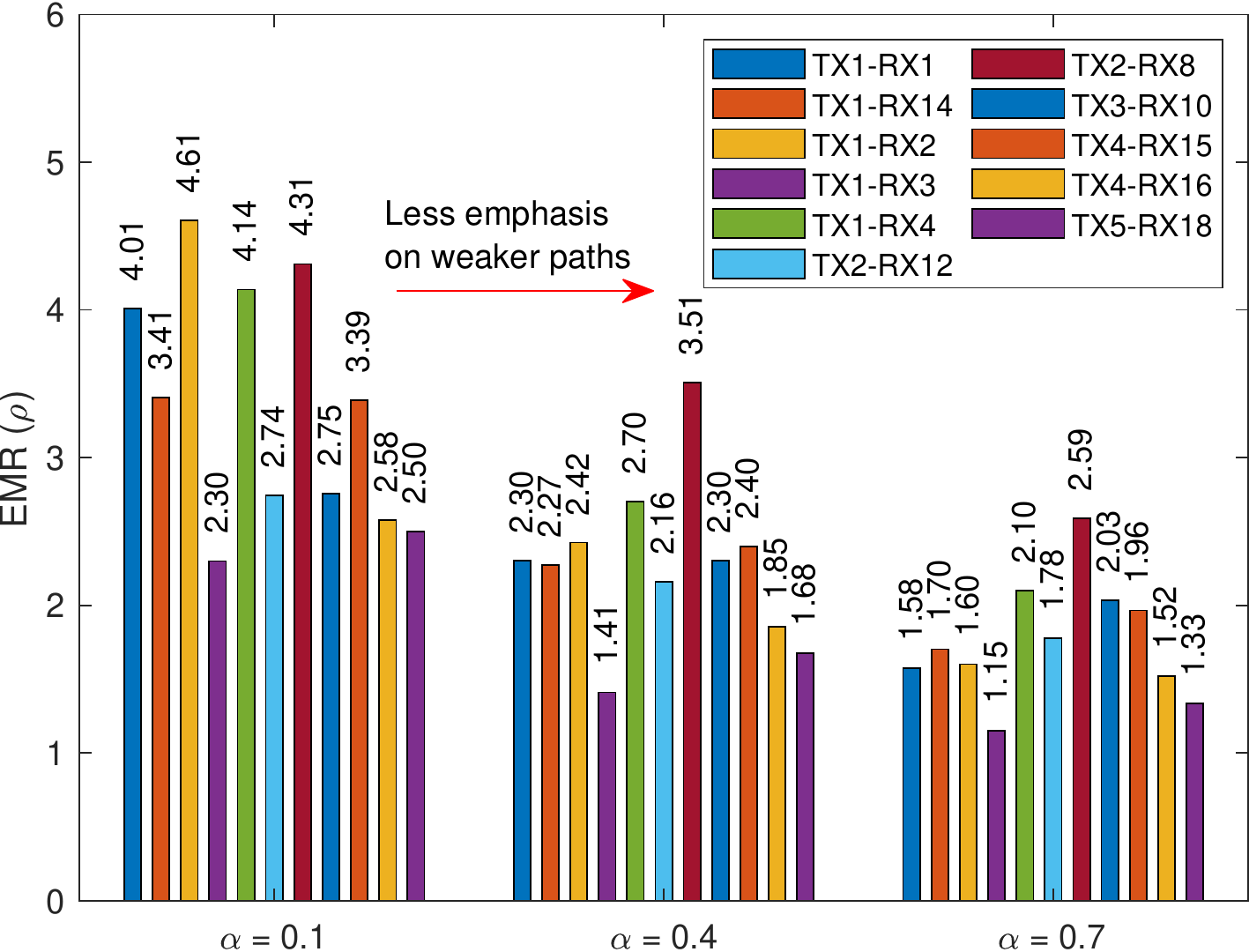}}
\hfill
\subfloat[]{\includegraphics[width=0.49\columnwidth]{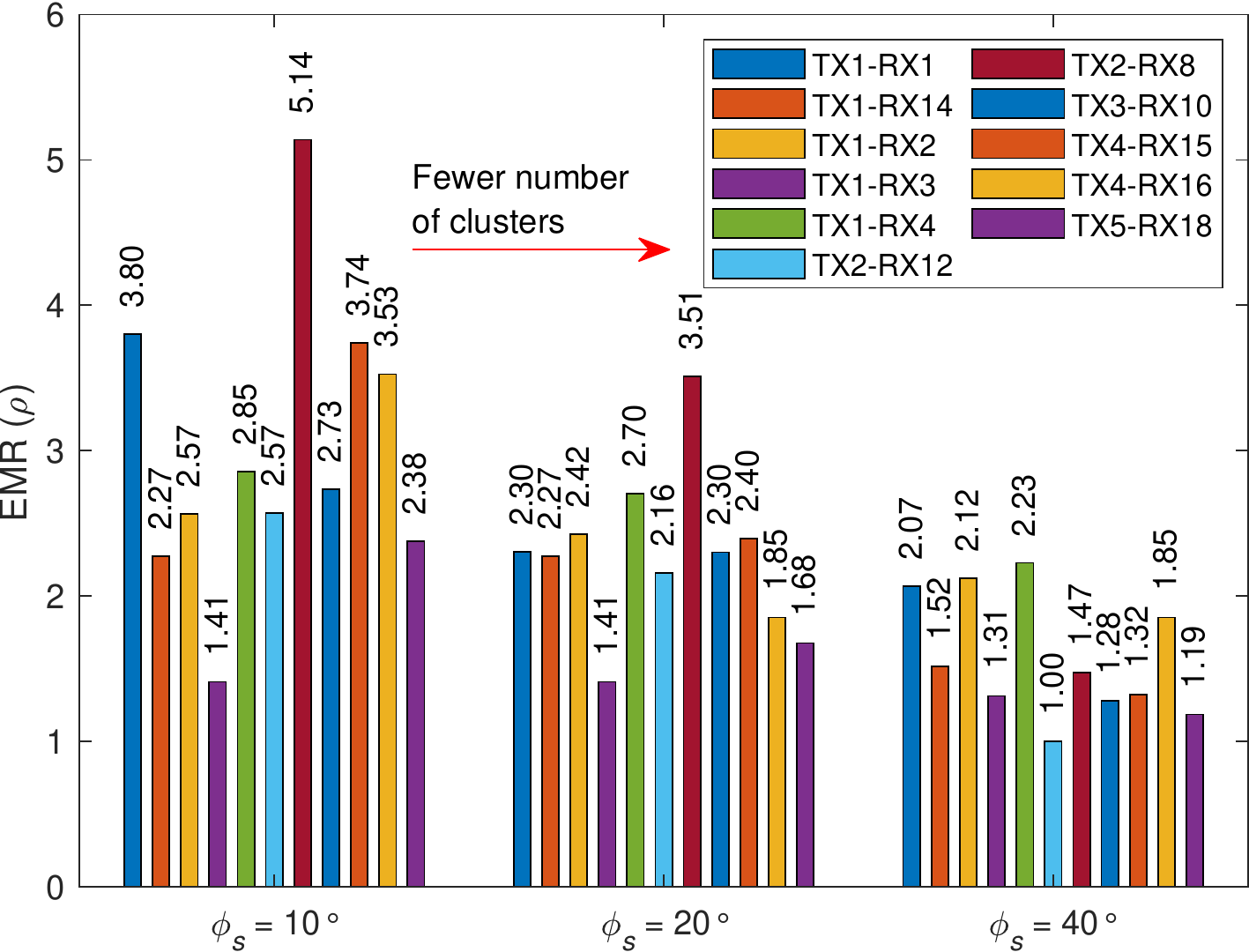}}
\vfill
\subfloat[]{\includegraphics[width=0.49\columnwidth]{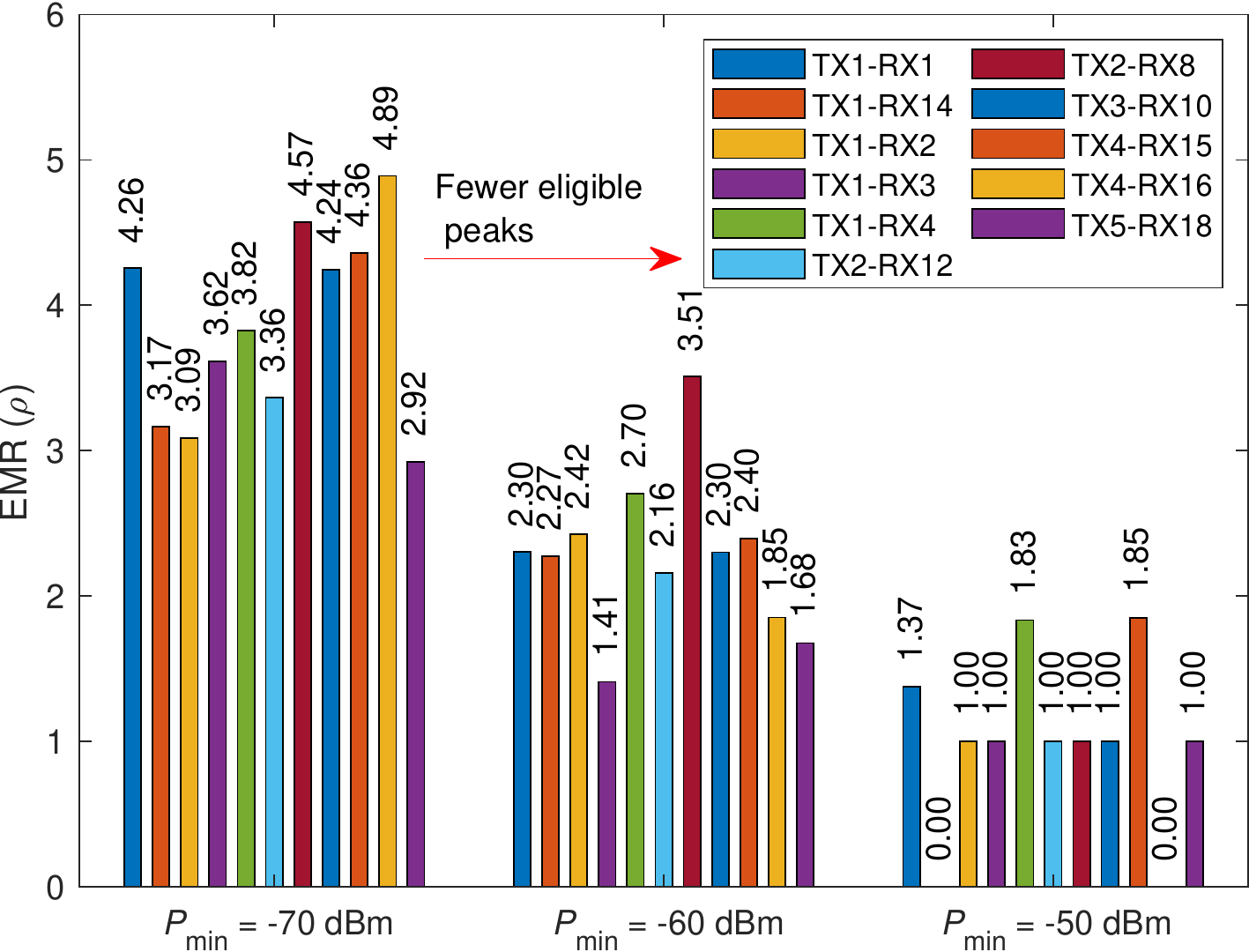}}
\caption{EMR for the LOS measurements for (a) varying $\alpha$ ($\phi_s=20^\circ$, $P_{{\min}}=-60$~dBm), (b) varying $\phi_s$ ($\alpha=0.4$, $P_{{\min}}=-60~\textrm{dBm}$), and (c) varying $P_{{\min}}$ ($\alpha=0.4$, $\phi_s=20^\circ$).}
\label{fig:rho_vs_parameters}
\vspace{-4mm}
\end{figure}

We evaluate the EMR metric for different $\alpha$, $\phi_s$, and $P_{\min}$ values. We change the value of one of the parameters while keeping the others fixed and repeat this procedure for all the parameters. Fig.~\ref{fig:rho_vs_parameters}(a) shows the change in the EMR with $\alpha$. As $\alpha$ is increased, weights of the weaker paths in~\eqref{eq:scam} decrease and hence the EMR decreases. The EMR attains the minimum value for the measurement at TX1-RX3 for all $\alpha$ values, i.e., 2.30, 1.41, and 1.15 for $\alpha=0.1$, $\alpha=0.4$, and $\alpha=0.7$, respectively. This result can be attributed to the geometry of the environment at these particular TX-RX locations which allows only the LOS path and reflections from a glass window. Thus having an EMR close to 1 means there is only one or two paths (other than the strongest path) with relatively low power for that TX-RX location and $\phi_s$ value. TX1-RX3 and TX2-RX12 both have the same number of clusters (see Fig.~\ref{fig:numClusters}); however, the EMR is notably larger for the latter. Moreover, as $\alpha$ increases, the rate of decrease in the EMR for TX1-RX3 is higher than in TX2-RX12. From this observation, it can be inferred that the alternate paths for TX2-RX12 have relatively higher power and hence they are more valuable. 

\begin{figure}[t!]
\centering
\subfloat[]{\includegraphics[trim=0.8cm 0cm 0cm 0.1cm,clip,width=0.49\columnwidth]{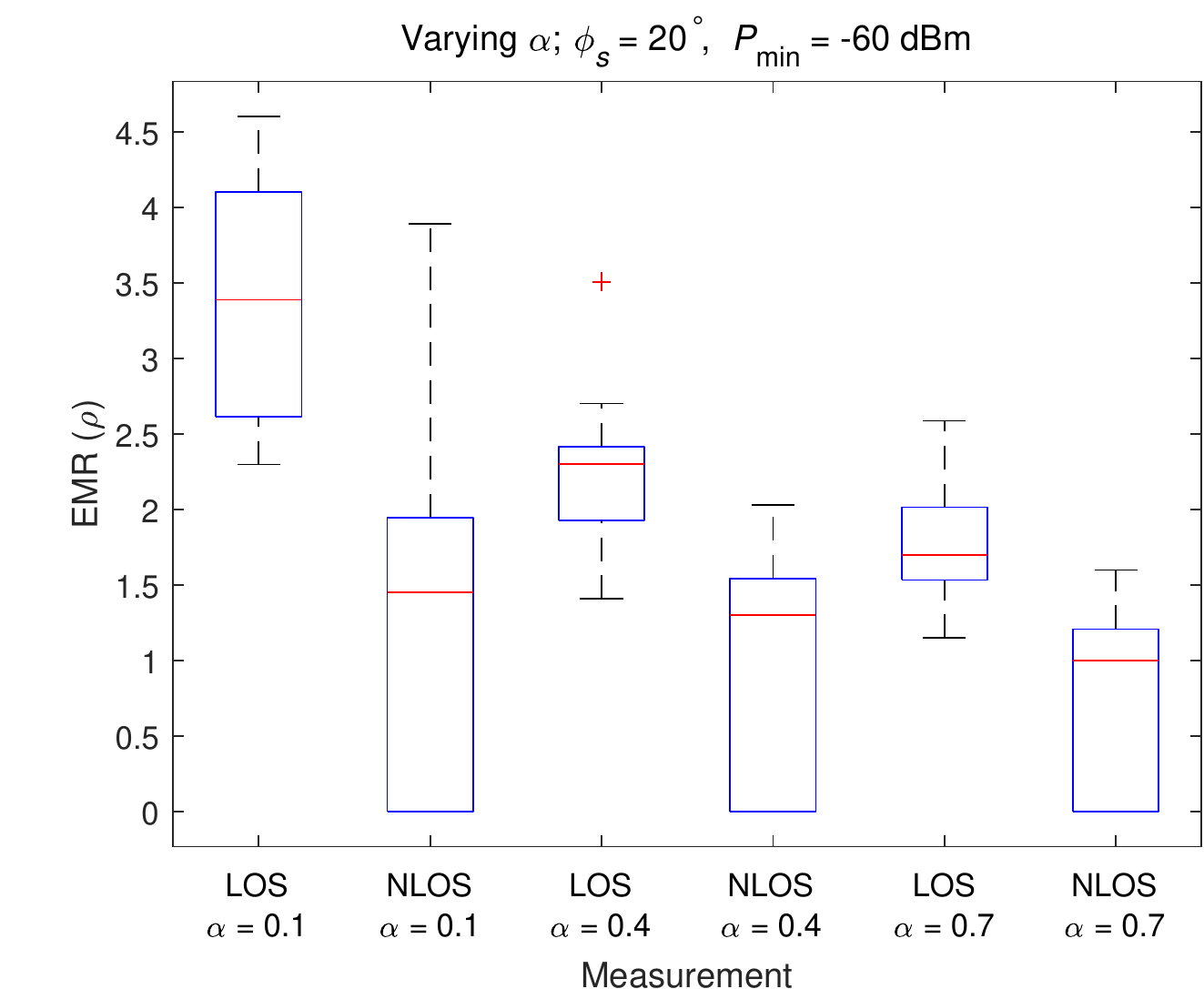}}
\hfill
\subfloat[]{\includegraphics[trim=0.8cm 0cm 0cm 0.1cm,clip,width=0.49\columnwidth]{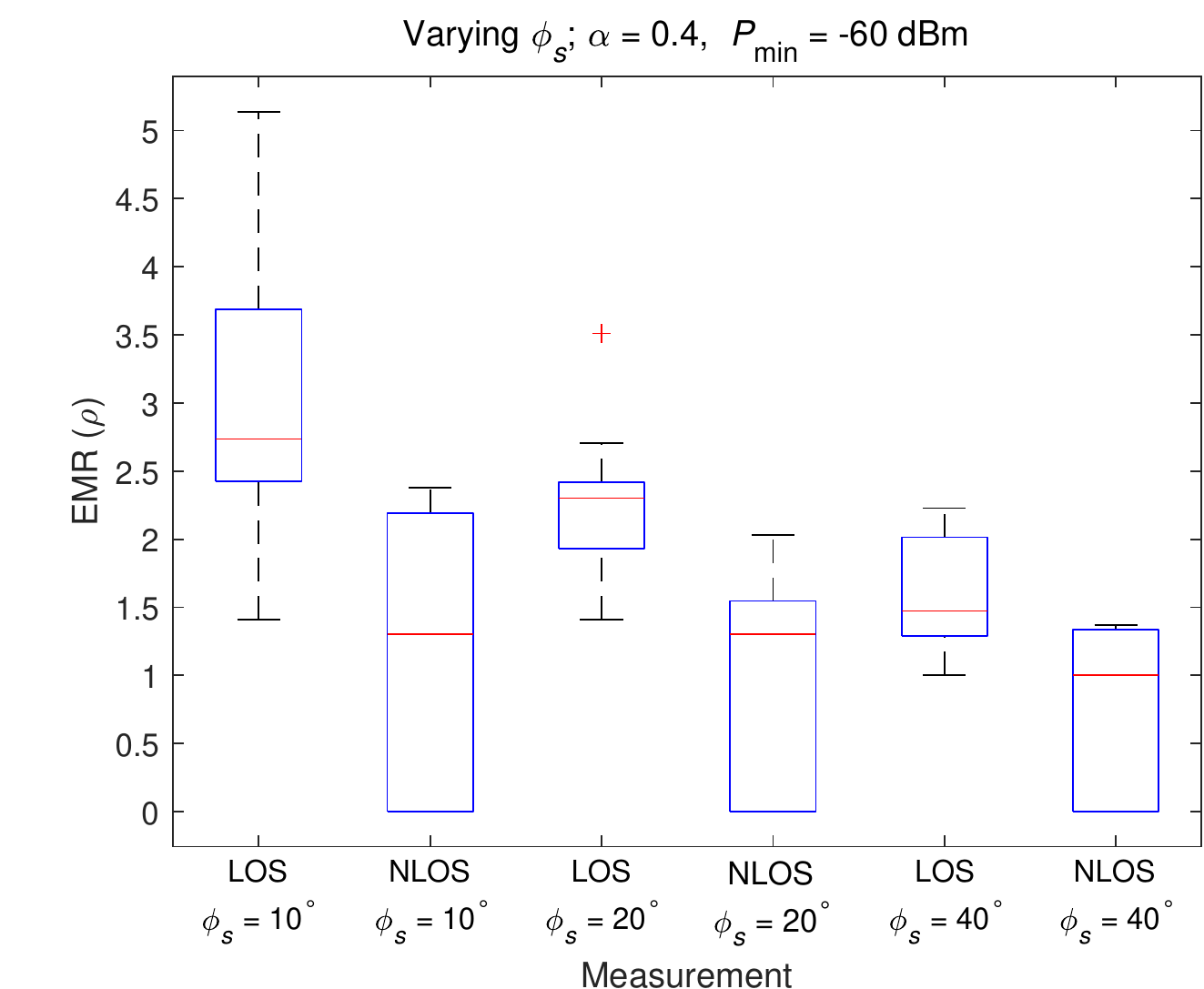}}
\vfill
\subfloat[]{\includegraphics[trim=1.0cm 0cm 0.05cm 0.1cm,clip,width=0.49\columnwidth]{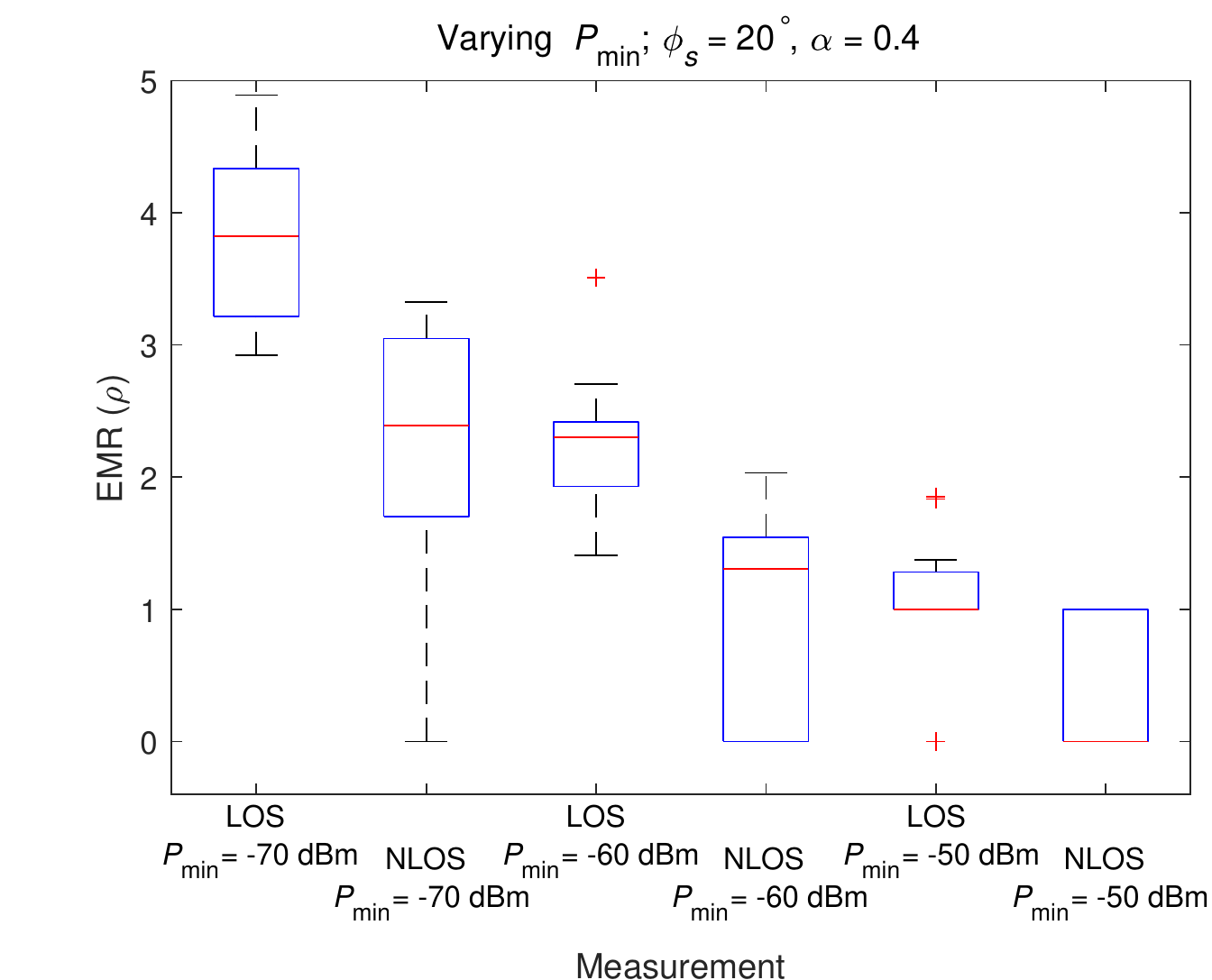}}
\vfill
\caption{Box plots of the EMR from LOS and NLOS measurements for (a) varying $\alpha$ ($\phi_s=20^\circ$, $P_{\min}=-60$~dBm), (b) varying $\phi_s$ ($\alpha=0.4$, $P_{\min}=-60~\textrm{dBm}$), and (c) varying $P_{\min}$ ($\alpha=0.4$, $\phi_s=20^\circ$).}
\label{fig:boxplots}
\vspace{-5mm}
\end{figure}





Fig.~\ref{fig:rho_vs_parameters}(b) shows the impact of $\phi_s$ on the metric. Since larger $\phi_s$ values make it more difficult to satisfy the constraint in~\eqref{eq:stopp}, there are fewer number of clusters with the increase in $\phi_s$. So fewer terms add up while calculating the EMR metric, which, in turn, results in smaller EMR values. When $\phi_s=40^\circ$, the EMR is calculated to be 1 for TX2-RX12, which means that the spatial diversity of the paths is poor. That is, there is only one effective path, and the communication can easily be broken down if that path is blocked. 

Next, we investigate the relation between the EMR and $P_{\min}$ in Fig.~\ref{fig:rho_vs_parameters}(c). Since increasing $P_{\min}$ results in fewer eligible peaks that meet the minimum power requirements, the EMR decreases. When $P_{\min}$ is increased up to $-50$~dBm, we observe that, for two of the measurements (i.e., TX1-RX14 and TX4-RX16), there is not a single path that has a power level above $-50$~dBm, and thus the EMR becomes zero. In addition, for most of the remaining TX-RX locations, there is only one effective path. 

To characterize the whole measurement environment, we give the box plots of the EMR metric in Fig.~\ref{fig:boxplots} for LOS and NLOS scenarios.
The horizontal lines indicate (from the top) the maximum, the 75th percentile, the median (also colored red), the 25th percentile, and the mimimum of the data. The plus signs represent the outliers.
We see that the EMR values for the NLOS measurements show the same trend with the change in the parameters as in the LOS measurements. In addition, the EMR values, as expected, are smaller in the NLOS measurements than the LOS measurements. Except for the case, where $P_{\min}$ is set to $-$70~dBm, the first quartile for the NLOS measurements is zero. This indicates that for some of the NLOS scenarios there is no link between the TX and the RX as mmWave frequencies are highly sensitive to blockages. However, we observe that, when $P_{\min}$ is below $-$50~dBm, the third quartile of the EMR for the NLOS measurements is above 1.5 for all the $\phi_s$ values considered. This implies that in most of the NLOS scenarios there are at least two beam directions that can be utilized for directional communication.

As indicated earlier, the EMR can also be used to quantify the effective multipath richness at a certain part of a measurement environment or to compare the candidate BS/AP locations in terms of their resilience to blockages. To demonstrate these use cases of the metric, in Table~V, we give the mean and standard deviation of the EMR calculated for all TX-RX location pairs on the second and fourth floor of the library. We also show the same statistics for two different TX locations on the same floors. It can be observed that the second floor provides a higher mean EMR than the fourth floor but the standard deviation is also higher for the former. A similar observation can be made for the TX locations TX1 and TX2, which are placed on the second floor. Even though TX2 provides a higher EMR on the average than TX1, the standard deviation of the EMR is also higher at TX2, making it a difficult decision to choose between the two locations. However, for the TX locations TX4 and TX5 on the fourth floor, the mean EMR is higher and the standard deviation of the EMR is smaller at TX4 than at TX5. Therefore, it can be concluded that TX4 is a more preferable location than TX5. Finally, we note that for a fair comparison between the EMR of different TX locations, one should perform several measurements at the same set of RX locations. Similarly, while comparing the EMR of different floors, to remove the bias in the EMR values, we suggest considering a comparable number of measurements for both LOS and NLOS scenarios on each floor.\looseness=-1

\begin{table}[t!]
\centering
\label{Tab:EMRbyFloor}
\caption {Mean and Standard Deviation of EMR for Different Floors and Different TX Locations on the Same Floor. TX1\&TX2 are on the Second Floor, and TX4\&TX5 are on the Fourth Floor.}
\renewcommand{\arraystretch}{1.3}
\begin{tabular}{lll||ll||ll}
\hline
EMR  & Floor 2 & Floor 4 & TX1  & TX2  & TX4  & TX5  \\ \hline
mean ($\rho$) & 1.81    & 1.66    & 1.72 & 1.98 & 2.08 & 1.24 \\
std ($\rho$)  & 1.06    & 0.55    & 0.89 & 1.5  & 0.28 & 0.38 \\ \hline
\end{tabular}
\vspace{-5mm}
\end{table}

\section{Conclusion}
\label{sec:Conclusion}
In this paper, we introduced a new metric, EMR, to assess the effective multipath richness in a communication channel. The proposed EMR metric takes into consideration the required power level and the spatial diversity of the paths. It also allows to adjust the relative importance of the paths that are not the primary choice. Using a single scalar value, EMR provides a measure of the suitability and resilience of a specific propagation environment for directional communications under blockage. The metric was evaluated for real channel measurements at 28~GHz in a library environment and shown to be useful in measuring the resilience of the propagation links to blockages. 
We believe that this metric will complement the other well-known propagation metrics, such as RMS-DS and RMS-AS, in characterizing different mmWave propagation environments, and improving the deployment and operation of next-generation mmWave networks. 


\bibliography{IEEEabrv,references}
\bibliographystyle{IEEEtran}

\end{document}